\documentclass{aa}

\usepackage{array, multirow, graphicx,natbib,txfonts,color,tabularx}
\bibpunct{(}{)}{;}{a}{}{,}

\newcommand{\ang}{$\rm \AA$}
\newcommand{\Ha}{H$\alpha$}
\newcommand{\Hb}{H$\beta$}

\newcommand{\kms}{km\,s$^{-1}$}

\newcommand{\ergs}{erg\,s$^{-1}$}

\newcommand{\flux}{erg\,cm$^{-2}$\,s$^{-1}$}

\newcommand{\Heone}{He\,{\scriptsize I}}
\newcommand{\HetwoX}{He\,$\mbox{\scriptsize{II}}$}

\newcommand{\Htwo}{H\,{\scriptsize II}}
\newcommand{\Othree}{$\left[\mbox{O \scriptsize{III}}\right]$}
\newcommand{\Oone}{$\left[\mbox{O \scriptsize{I}}\right]$}
\newcommand{\Ntwo}{$\left[\mbox{N \scriptsize{II}}\right]$}
\newcommand{\Stwo}{$\left[\mbox{S \scriptsize{II}}\right]$}
\newcommand{\Hetwo}{He\,{\scriptsize II}\,$\lambda4686$}

\newcommand{\OthreeB}{$\left[\mbox{O \scriptsize{III}}\right]$\,$\lambda5007$}
\newcommand{\OthreeC}{$\left[\mbox{O \scriptsize{III}}\right]$\,$\lambda\lambda4959,5007$}
\newcommand{\OoneA}{$\left[\mbox{O \scriptsize{I}}\right]$\,$\lambda6300$}

\newcommand{\OoneC}{$\left[\mbox{O \scriptsize{I}}\right]$\,$\lambda\lambda6300,6363$}

\newcommand{\NtwoB}{$\left[\mbox{N \scriptsize{II}}\right]$\,$\lambda6583$}
\newcommand{\NtwoC}{$\left[\mbox{N \scriptsize{II}}\right]$\,$\lambda\lambda6548,5683$}
\newcommand{\StwoA}{$\left[\mbox{S \scriptsize{II}}\right]$\,$\lambda6716$}
\newcommand{\StwoB}{$\left[\mbox{S \scriptsize{II}}\right]$\,$\lambda6731$}
\newcommand{\StwoC}{$\left[\mbox{S \scriptsize{II}}\right]$\,$\lambda\lambda6716,6731$}

\begin{document}


\title{A VLT VIMOS IFU study of the ionisation nebula surrounding the Supersoft X-ray Source CAL 83 }

\author{Pieter Gruyters \inst{1,2} \and Katrina Exter \inst{1} \and Timothy P. Roberts \inst{3} \and Saul Rappaport \inst{4}}

\offprints{pieter.gruyters@gmail.com}

\institute{Instituut voor Sterrenkunde, K.U.Leuven, Celestijnenlaan 200D,
B-3001 Leuven, Belgium \and Department of Physics and Astronomy, Division of Astronomy and Space Physics, Uppsala University, Box 516, 75120 Uppsala, Sweden \and Department of Physics, Durham University, South Road, Durham DH1 3LE, UK \and Department of Physics and Kavli Institute for Astrophysics and Space Research, Massachusetts Institute of Technology, Cambridge, MA 02139, USA}

\date{Received  / Accepted}

\authorrunning{P. Gruyters}
\titlerunning{The ionised nebula surrounding CAL 83}

\abstract
{CAL\,83 is a prototype of the class of Super Soft X-ray Sources (SXS). It is a binary consisting of a low mass secondary that is transferring mass onto a white dwarf primary and is the only known SXS surrounded by an ionisation nebula, made up of the interstellar medium (ISM) ionised by the source itself. We study this nebula using integral field spectroscopy. }
{The study of ionised material can inform us about the source that is responsible for the ionisation, in a way that is complementary to studying the source directly. Since CAL\,83 is the only SXS known with an ionisation nebula, we have an opportunity to see if such studies are as useful for SXSs as they have been for other X-ray ionised nebulae. We can use these data to compare to models of how CAL\,83 should ionise its surroundings, based on what we know about the source emission spectrum and the physical conditions of the surrounding ISM.}
{With the VIMOS integral field spectrograph we obtained spectra over a $25 \times 25$\arcsec field of view, encompassing one quarter of the nebula. {\bf Emission line maps -- H\,{\sc i}, \Hetwo, \OthreeC, \NtwoC, and \StwoC\ -- }are produced in order to study the morphology of the ionised gas. We include CAL\,83 on diagrams of various diagnostic ion ratios to compare it to other X-ray ionised sources. Finally we computed some simple models of the ionised gas around CAL\,83 and compare the predicted to the observed spectra. }
{CAL\,83 appears to have a fairly standard ionisation nebula as far as the morphology goes: the edges where H is recombining are strong in the low stage ionisation lines and the central, clumpy regions are stronger in the higher stage ionisation lines. But the He\,{\sc ii} emission is unusual in being confined to one side of CAL\,83 rather than being homogeneously distributed as with the other ions. We model the CAL\,83 nebula with {\sc cloudy}\,\normalfont {\bf using model parameters for SXSs found in the literature}. The He\,{\sc ii} emission does not fit in with model predictions; in fact none of the models is able to fit the observed spectrum very well.}
{The spectral line images of the region surrounding CAL\,83 are revealing and instructive. However, more work on modelling the spectrum of the ionised gas is necessary, and especially for the high-ionisation level emission from CAL\,83. In particular, we wish to know if the He\,{\sc ii} emission {\bf and the other nebular lines are powered by the same ionising source.}}


\maketitle


\section{Introduction}\label{sect:intro}

Between 1979 and 1981, the Columbia Astrophysics Laboratory (CAL) carried out a systematic soft X-ray survey of the large Magellanic cloud (LMC) using the \textit{Einstein} \textrm{Observatory}. This resulted in the detection of 97 X-ray sources \citep*{Long1981}. Some of the detected sources are characterised by an unusually soft spectrum in which little or no radiation at energies above $\sim0.5$ keV is detected. Source no. 83, being one of a small number of such sources, became known as CAL\,83 (also known as LHG83, $RX J0543.7$-6822, and $1E0543.8$-6823) and it is now regarded as the prototype of the class of supersoft X-ray sources (SXSs). In 1990, the first X-ray all-sky survey was performed with \textrm{ROSAT} and many more similar sources were found in the Galaxy, LMC, and M31 (\citealt*{Kahabka1997}, \citealt{Parmar1998}). These sources are characterised by a luminosity of $\sim$\,$10^{37-38}$ \ergs\ and effective temperatures in the range of $\sim$\,$2-6\times10^5$ K ($kT\simeq17$--50 eV) \citep*{Rappaport1994}. 

Van den Heuvel et al.\ (1992) were the first to propose a model for SXSs: an accreting white dwarf of mass 0.7--1.2$M_{\odot}$ accompanied by a normal star of mass 1.5--2$M_{\odot}$. According to their model the supersoft X-ray emission and high luminosities are a result of steady nuclear burning of hydrogen accreted onto the white dwarf. The mass transfer from the main-sequence star to the white dwarf occurs on a thermal timescale via Roche-lobe overflow, with rates of $\sim1-4\times10^{-7}M_{\odot}$yr$^{-1}$. The dominant optical light source is the white dwarf, which, with its accretion disc, completely outshines the donor star. The nuclear burning on the surface of the white dwarf allows it to retain the accreted mass, making it a possible progenitor of accretion-induced collapse \citep{Heuvel1992} or Type Ia supernovae \citep{Rappaport1994, Nelson1996, Hachisu1996}.

CAL\,83 has a time-averaged luminosity $L_{X}$\,(0.15--4.5\,keV) of $3.2\times10^{37}$ \ergs\ and is identified with a variable B$\sim$16.8 star with a blue continuum and strong narrow ($\sim$$305$ \kms), variable \Hetwo\ emission \citep{Cowley1998}. It is, in fact, a mass-transferring binary, with a secondary mass of $\sim0.5M_{\odot}$ \citep{Cowley1998} and a white dwarf primary mass of $1.3\pm0.3 M_{\odot}$ \citep{Alcock1997,Lanz2005}. The orbital period of the system is 1.04 days \citep{Smale1988}.
While CAL\,83 is one of a dozen luminous, extremely soft SXSs now known in our Galaxy and the Magellanic Clouds, so far it is the only one known to have an ionisation nebula. Its inhomogeneous nebula has bright emission extended over 37\arcsec, with fainter emission extending out beyond twice that, and a total mass of $\sim150M_{\odot}$ (\citealt*{Remillard1995} = RRM95); obviously the nebula is ionised interstellar medium (ISM), rather than being intimately related to CAL\,83 itself. 

SXSs emit copious quantities of photons in the range 20--200 eV, hence the radiation can ionise any gas surrounding them, creating an ionisation nebula. We expect these ionisation nebulae to be distinct from classic H\,\scriptsize{II}\normalsize\ regions where the ionisation is the result of photo-ionisation by massive O-type stars by the absorption of higher energy photons while the lower energy photons escape. In comparison, in the case of the SXSs it is the other way around: the lower energy photons do the ionising while the higher energy photons escape (RRM95). Models for SXS nebulae by \citet{Chiang1994} predict that these should be distinct from other astrophysical nebulae, in particular, \Othree\ and \Hetwo\ should be far brighter than in classic \Htwo\ regions, and the radial gradients of these and other lines much more gradual.
These models have not yet been tested on real data of SXSs, and CAL\,83 presents a good opportunity to do so.

Using the VLT VIMOS \citep{Lefevre2003} integral field unit (IFU) we have observed one field around CAL\,83, obtaining spatial and spectral information about its surrounding ISM. In this paper we report on our findings, presenting emission line flux maps, spectra, line ratio maps, and a comparison of our results to models of ionised ISM created with the {\sc cloudy} code \citep{Ferland1998}.



\section{Observations and data reduction} \label{sect:data}
\subsection{Observations}
CAL\,83 was observed at the VLT of ESO using the VIMOS instrument in its IFU mode. We used the small field-of-view (FoV) mode, resulting in a coverage of 27\arcsec $\times$ 27\arcsec, covered by 1600 spatial pixels (\emph{spaxels}) at a spatial sampling of 0.67\arcsec per spaxel. The data were taken on 2005 Oct.\ 28 and Dec.\ 1. The spatial sampling is contiguous with a dead space between spaxels of less than 10 per cent of the spaxel separation \citep{Zanichelli2005}. We used the high resolution blue grism (HRblue, $\sim$0.54\ang\ pixel$^{-1}$), covering a spectral range of 4150--6200\ang\ and the high resolution orange grism (HRorange, $\sim$0.62\ang\ pixel$^{-1}$) covering 5250--7400\ang. The observing log can be found in Table \ref{Tab:obslog}. We have 2 exposures per grism, and there is a slight offset in the pointing for some of the observations. For the flux calibration we made use of two standard star observations, one for each resolution. As is standard ESO practice, these are not (necessarily) observed on the same night as the astronomical target. Table \ref{Tab:obslog} also includes the details of the data from which we derived our sky spectrum (Sec. 2.2.2).

\begin{table*}[ht]
\caption[The VIMOS IFU observing log.]{\label{Tab:obslog}
The VIMOS IFU observing log.}
\begin{center}
\begin{tabular}{lcccccc} \hline\hline\rule[0mm]{0mm}{3mm}
Name & Observation ID & Date & Grism & Exp. time (s) & RA & Dec \\
\hline  
Blue1 & 206671 & 28/10/2005 & HR blue & $1\times1320$ & 05 45 33.34 & -68 22 19.2 \\
Orange1 & 206675 & 01/12/2005 & HR orange & $1\times1320$ & 05 45 33.35 & -68 22 19.4 \\
Orange2 & 206676 & 01/12/2005 & HR orange & $1\times1320$ & 05 45 33.11 & -68 22 18.0 \\
Blue2 & 206677 & 01/12/2005 & HR blue & $1\times1320$ & 05 45 33.11 & -68 22 18.0 \\
\hline 
Standard 1 & 200152644 & 28/10/2005 & HR blue & $1\times123$ & 00 23 48.97 & -27 51 43.1 \\
Standard 2 & 200152643 & 28/10/2005 & HR orange & $1\times154$ & 00 23 48.97 & -27 51 43.1\\ 
Sky & 200152515 & 12/10/2005 & HR orange & $1\times123$ & 23 20 42.56 & +05 09 47.9 \\
\hline\hline
\end{tabular}
\end{center}
\end{table*}

\subsection{Data reduction}
The data were reduced and flux calibrated using the ESO pipeline \textsc{Gasgano}\footnote{http://www.eso.org/sci/software/gasgano}, which allows the user to organise calibration files and run pipeline tasks. The reduction involves three main tasks: (1) \emph{vmbias} which creates the master bias frame; (2) \emph{vmifucalib} which determines the spectral extraction mask, wavelength calibration and the relative fibre transmission corrections; and (3) \emph{vmifustandard} for producing the flux response curves from a summed spectrophotometric standard star spectrum. Finally, the output of these tasks are fed into \emph{vmifuscience} which extracts the bias-subtracted, wavelength- and flux-calibrated, and relative fibre transmission-corrected science spectra. The FoV of the VIMOS IFU is split into four quadrants, and the light from each quadrant goes to a separate CCD, therefore the data processing is performed separately on each quadrant, creating four fully calibrated frames per science exposure. 

Before doing the flux calibration it was necessary to step out of the pipeline to improve on the fibre transmission correction. Looking at the (Gaussian-profile fit) flux of the sky line at 5577\,\AA\ in each spaxel for each quadrant separately, clear systematic intensity variations of the order 20--30\%\ were found: the ``spectrum" of flux versus spaxel number showed the same pattern for all the frames of the same grism. For each frame and each quadrant of each grism, we measured the deviation of each spaxel with respect to the overall median of the individual quadrants. This value was then divided out of the science frames, reducing the variations to a random scatter of 5--10\%.

Finally, it was necessary to apply multiplicative corrections to the spectra of the four quadrants so that the quadrant-mean 5577\,\AA\ flux was the same for all quadrants. These rescaling factors are 1.3, 0.9, 0.8 and 0.9 respectively for quadrants 1 to 4, with scatter between the different observations.

\subsubsection{Combining data}
The final data cube for each grism was produced by averaging the two exposures of each. The slight pointing offsets between the frames had to be corrected for before the frames could be combined. We used the IRAF routine {\sl drizzle} \citep*{Mutchler1997}, which was developed for \textit{Hubble} \textrm{Space Telescope} (HST) images, to do this. For each wavelength of our spectral range, we extracted the 2D (flux) image, and spatially resampled the one frame to match the world coordinate system (WCS) of the other frame, for the blue and orange data separately. The offsets to apply were determined from the three stars that are in our FoV, by measuring their position to a precision of 1/2 spaxel in either direction using continuum images constructed from our data cubes and comparing between the images. The offsets are the same for all wavelengths (there is no evidence for differential atmospheric refraction in our data). Offsets are of the order 1.5--2.5 spaxels. As a result of these offsets, the final maps are slightly smaller than the IFU FoV, being 25.5\arcsec$\times$\,25.5\arcsec. For the subsequent analyses we placed both cubes on the same artificial WCS grid (but with the correct spaxel relative positions and sizes) -- the slight offset in pointing between the blue and orange cubes is hence noticeable in the figures in this paper. 

\subsubsection{Sky subtraction}
\begin{figure}
\begin{center}
\includegraphics[width=0.95\columnwidth]{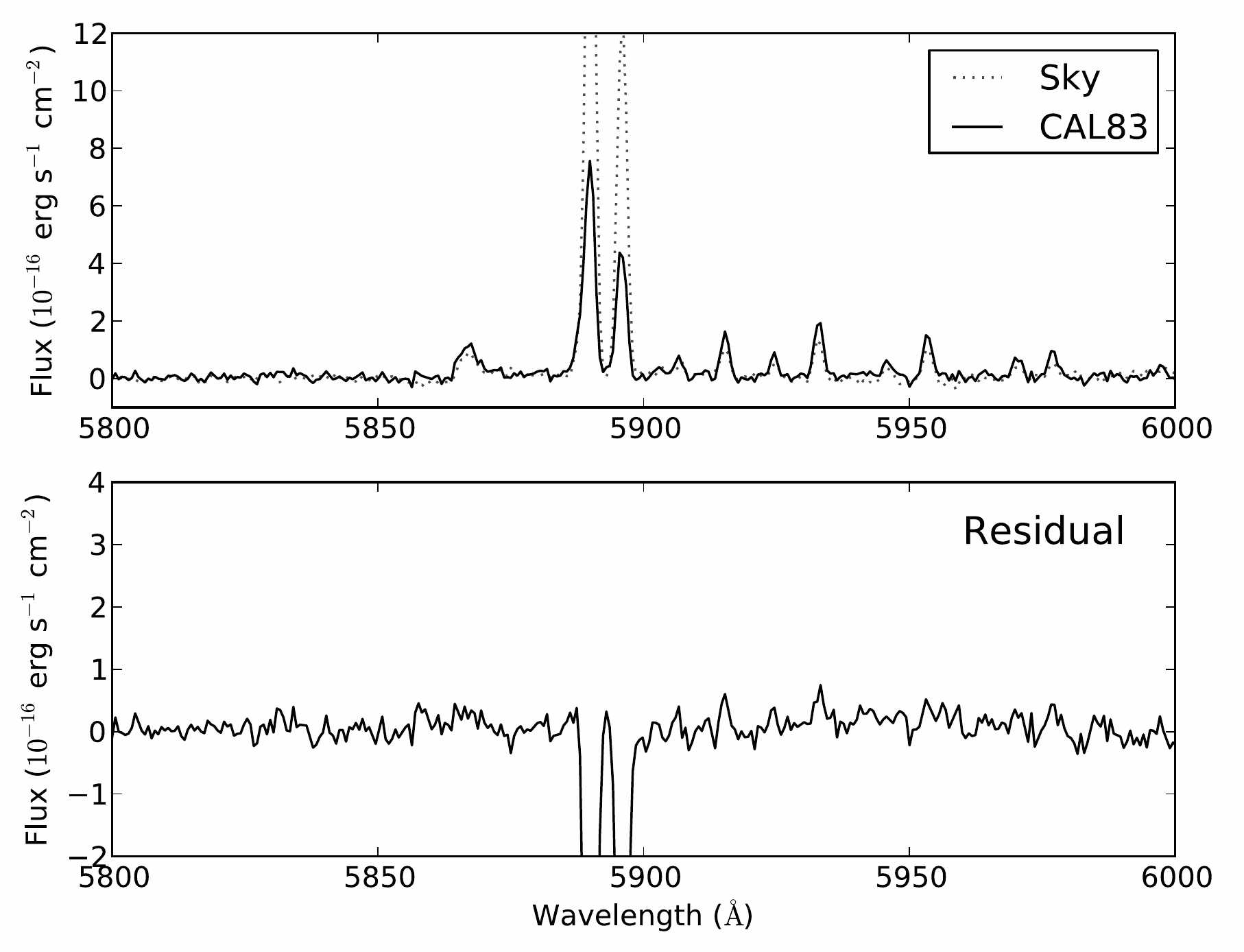}
\includegraphics[width=0.95\columnwidth]{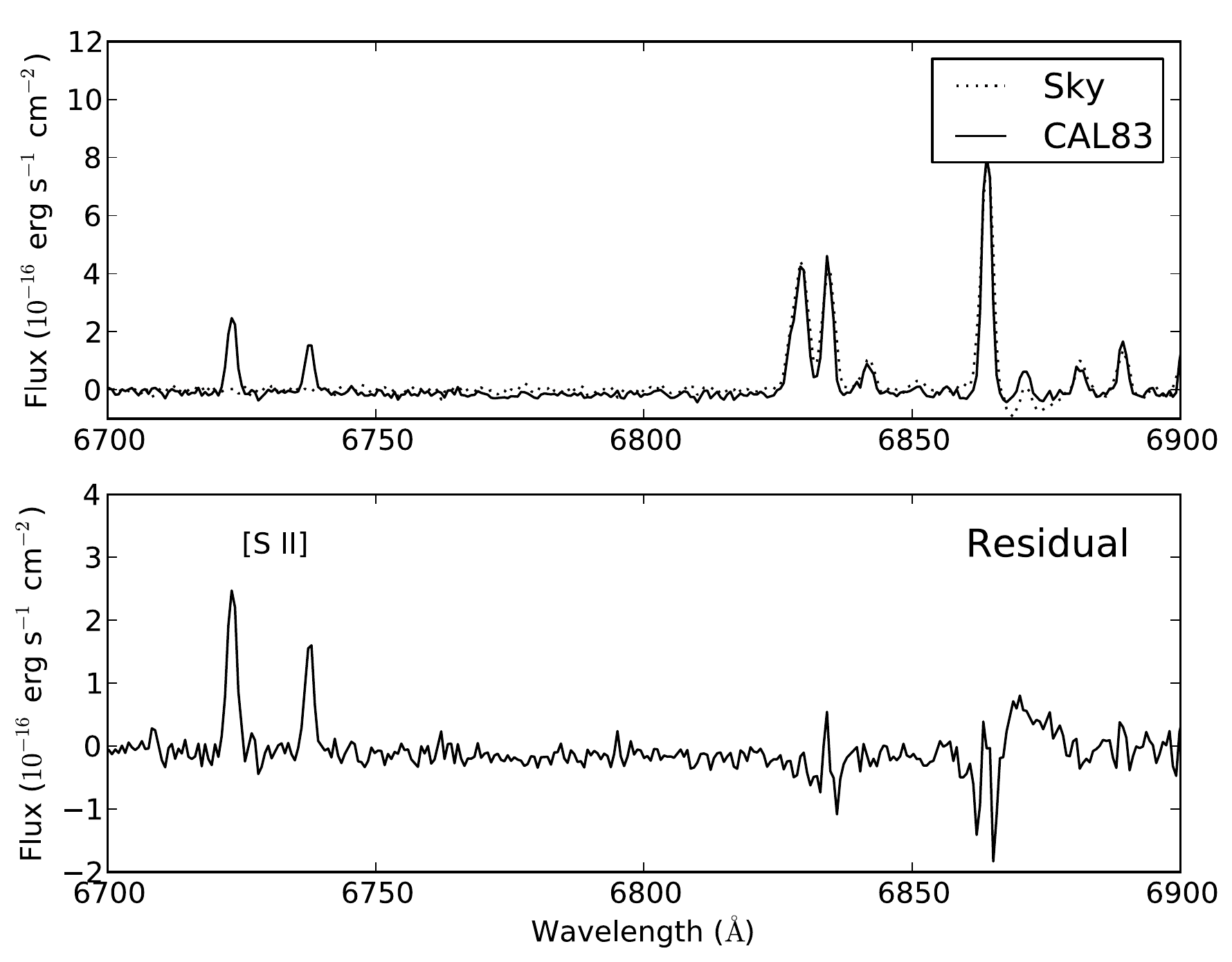}
\caption{Sky subtraction. Shown is the spectrum of a single spaxel of CAL 83 with the scaled telluric spectrum overplotted. Below each plot are the residuals. On the \textit{ top} one notices the difference in the flux of the brightest sky lines with the rest (see text). On the \textit{ bottom} one can see S-shaped residuals which are a result of spectral PSF differences (see text). }\label{Fig:sky}
\end{center}
\end{figure}

Sky subtraction can be a problem with the VIMOS IFU, as there are no sky-dedicated fibers. In our FoV there are no spaxels with only telluric emission. Therefore we looked in the ESO archive for long observations of standard stars taken with our IFU set-up and close in time to our observations. These frames are listed in Table 1. The FoV of these frames contain only one star, centred on each quadrant. We reduce the data through the pipeline, including flux calibration. The spaxels clear of stellar emission were then extracted. Ideally we would have one sky spectrum per quadrant, or even per spaxel, as there are slight variations in the spectral PSF with position on the CCD. However, lack of signal-to-noise ratio meant we had to sum all the sky data into a single template sky spectrum. To subtract this template from our data it was necessary to scale the template sky spectrum. We measured the line fluxes from the template and from each spaxel in our CAL\,83 frames. There is no variation, outside of noise, in the sky fluxes from CAL\,83 with spaxel (just as one would expect) and hence we scaled the template sky to the median of the whole CAL\,83 field. The sky lines chosen for scaling were not the brightest -- $\lambda\lambda$5577, 5890, 5896 (and secondarily 6300, which can also have some Galactic contribution) -- as this resulted in an under-subtraction of all other sky line fluxes. This may be due to variation in the telluric spectrum with time. Instead we scaled to a dozen or so mid-bright lines. The results of our sky subtraction are shown in Fig. \ref{Fig:sky}. In the plot showing the reddest spectral range, S-shaped residuals can be seen: this is a consequence of variations to the spectral PSF with position on the CCD which we could not correct for. 

We carried out the sky subtraction only for the orange data. For the blue data the telluric spectrum is less of a problem as there are no sky lines that can interfere with our CAL\,83 spectral lines; and we are not interested in the continuum level, only the emission line fluxes. The only potential contamination can be at He\,{\sc i}\,5876\,\AA\ and around 5200\,\AA\ (where lines of [N\,{\sc i}] may be found). Careful inspection of our data (in cube form, in row-stacked spectrum form, and comparison to a blue sky spectrum) shows that there is no He\,{\sc i} (redshifted: 5881\,\AA) and no [N {\sc i}] ($\sim$5205\,\AA) emission above telluric. 

\subsection{Emission line fitting and map making}
Our final frames contain 1600 spectra each. To fit spectra of this quantity we adopted an IDL-based semi-automated fitting procedure called \texttt{PAN} (Peak ANalysis; \citealt{Dimeo2005}), modified by \citet{Westmoquette2007a} to accept row-stacked-spectrum-format FITS files. The user chooses a spectrum in the stack, a spectral region to fit, and defines the models -- usually a polynomial to the continuum and one to a few Gaussian profiles to emission lines -- and the software proceeds to fit all spectra in the input frame. The Levenberg-Marquardt technique is implemented to solve the least-squares minimisation problem during the fitting. 

For PAN to produce its best result a realistic error spectrum must be attached to each source spectrum. These error spectra were created by us in the following way: the raw science frames (with signal values of ADU) were converted to (assumed Poissonian) percentage error frames, adjusted and combined in the same way that the science frames had been, and then multiplied into the final CAL\,83 frames to create error frames in physical units. These were attached to the CAL\,83 FITS files and read into PAN. 

PAN produces residual spectra (calculated as $r_i = (y_{i}^{fit}-y_{i}^{data})/\sigma_i$, where ${\sigma_i}$ are the data uncertainties), and the model results (for the Gaussians these are the central wavelength, integrated flux and FWHM, plus their errors), the spaxel numbers, and the $\chi^2$ values are output to a file. Inspection of the residual spectra, the errors and $\chi^2$ values, allows one to discard doubtful results. 

These fits were later to be turned into emission line maps. To do this one requires only the table (provided by ESO) that relates the spaxel number to its relative sky coordinates. Constructing the maps, displaying as greyscale or contour, and plotting maps on top of each other (contour over greyscale and accounting for shifts in the WCS) were done from the PAN-output files with {\emph Jython} scripts and using the tools provided by HIPE, the \textit{Herschel} Interactive Processing Environment \citep{ott2010}.

Note that the data were not dereddened before being fit in PAN.

\subsection{Dereddening}
To correct for reddening we used the Galactic reddening law of \citet{Howarth1983} with the $c(\mbox{\Hb})$ values derived from our observed Galactic \Ha/\Hb\ line ratios in conjunction with the theoretical Case B ratio from \citet{Osterbrock1989}, which is 2.85 for a temperature of $10,000$\,K. 
We calculated a $c(\mbox{\Hb})$ value for all spaxels in the FoV using the Galactic \Ha/\Hb\ ratio. We then dereddened our observed LMC \Ha\ and \Hb\ fluxes (the LMC redshift separates them from the Galactic lines) and calculated the LMC $c(\mbox{\Hb})$ values. This, however, resulted in a FoV with negative $c(\mbox{\Hb})$ values. As the map of our Galactic $c(\mbox{\Hb})$ values is flat, we opted to use one median Galactic $c(\mbox{\Hb})$ value for the whole FoV to deredden the data (i.e. we assume all the reddening is Galactic). The median ratio for \Ha/\Hb\ derived from all the good signal-to-noise spaxels in our FoV is $3.8\pm0.5$, corresponding to $c(\mbox{\Hb}) = 0.38_{-0.20}^{+0.17}$, which by following \citet{Kaler1985} and using $E(B-V)=c(\mbox{\Hb})*(0.61+(0.024*c(\mbox{\Hb})))$, is a value of $E(B-V)=0.23$. This value is however fairly high and cannot be reconciled with values found in the literature, i.e., the $E(B-V)$ value found by \citet{Kovacs2000} is 0.13, \citet*{Oestreicher1996} deduced a range of $E(B-V)$ of (0--0.15) for the Galactic foreground extinction toward the LMC, and RRM95 adopted $E(B-V) = 0.1$ in their paper about the nebula surrounding CAL\,83. However if we adopt the lower limit for the $c(\mbox{\Hb})$ value we find a value of $E(B-V)=0.11$, in perfect agreement with the literature. Hence we adopt $c(\mbox{\Hb}) = 0.18$ to deredden our fluxes. We repeat, we chose this lower limit mainly to make comparison to previous CAL 83 data more straightforward, and because the lower limit is within the standard deviation (scatter) in our field of view.

All values quoted henceforth (in tables, images and plots) are the dereddened values. The spectra presented in the paper are not corrected for extinction.



\section{Results} 
\label{sect:analysis}

With the aid of emission line maps we can discuss what the CAL\,83 nebula looks like. We will compare these emission maps to previous observations of the nebula. We have also extracted the global spectrum of the nebula and of the CAL\,83 stellar source, and will present the latter and briefly compare the stellar He\,{\sc ii} line to what is already known of this feature.

\subsection{Global spectrum}
\label{sect:Global spectrum}

Fig.\,2 shows our spectrum of the ionisation nebula surrounding CAL\,83 taken from our 25.5\arcsec $\times$ 25.5\arcsec\ FoV, which corresponds to an actual size of $7.5\times7.5$\,pc at the distance of the LMC (55kpc; \citealt{Smale1988}). The stars (CAL\,83 itself and two field stars) have been excluded from this spectrum. The emission lines in the spectrum are labelled. Of particular interest is the detection of \Hetwo\ which has never been detected {\sl in the nebula} before. In addition to \Hetwo\ we detected the usual forbidden and Balmer lines. We also detected \OoneC\ and Balmer emission from our own Galaxy. From this spectrum, we measured the emission line fluxes using the ELF package in DIPSO \citep{Dipso}. The results of the Gaussian fits to the emission lines are given in Tables \ref{Tab:fluxes} and \ref{Tab:faint}. There we quote the integrated reddened and dereddened fluxes derived from all spaxels without stellar emission in them, i.e., only the ISM spectrum (summing 1213 and 1233 spaxels together respectively for the blue [4300--5100\,\AA] and orange [6300--6800\,\AA] emission lines). Along with the integrated fluxes we also present the average flux per $0.67\times0.67$\,\arcsec spaxel. The quoted errors in the tables include the flux and fitting errors. Reddening errors are not included, as is standard practice. To give a feeling for the order of magnitude of the reddening errors, for \Hb\ and \Ha, with reddening errors included the fluxes would be $(387\pm75)\times10^{-16}$\,erg\,cm$^{-2}$\,s$^{-1}$ and $(968\pm213)\times10^{-16}$\,erg\,cm$^{-2}$\,s$^{-1}$.

\begin{figure*}[htb]
\begin{center}
\label{Fig:spectra}
\includegraphics[width=1.7\columnwidth]{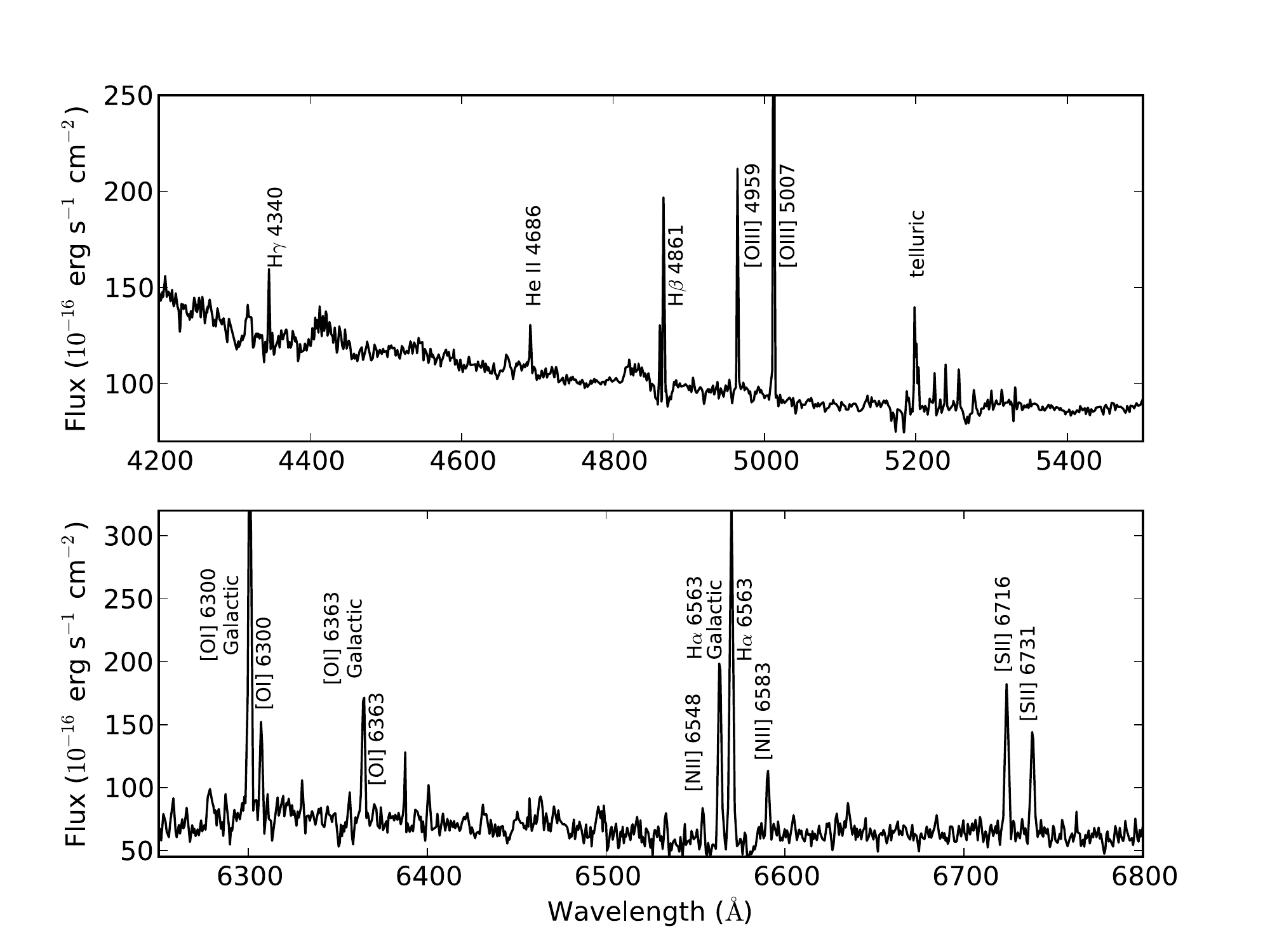}
\caption{Spectrum of the ionisation nebula of CAL\,83 taken from our 25.5\arcsec $\times$ 25.5\arcsec\ ($7.5\times7.5$ pc$^2$) FoV, excluding the contribution from the three stars (2 field stars and CAL\,83). The spectra were extracted from 1213 and 1233 spaxels for the blue and red part of the spectrum, so the area from which the spectra are extracted differs between the blue and the orange observations. {\sl Note that the spectra are not dereddened}.}
\end{center}
\end{figure*}

\begin{table*}[ht]
\caption[Emission line Fluxes of the total FoV]{\label{Tab:fluxes}
Emission line measurements for the summed spectrum of the CAL\,83 nebula. }
\begin{center}
\begin{tabular}{lcccccc} \hline\hline\rule[0mm]{0mm}{3mm}
Line \hspace{0.5cm} & \hspace{0.3cm}$\lambda_{air}$ \hspace{0.3cm} & $f(\lambda$) & \hspace{0.3cm}$F(\lambda)$\hspace{0.3cm} & \hspace{0.3cm}$I(\lambda)$\hspace{0.3cm} & no. spaxels & \hspace{0.3cm}$I(\lambda)_{\mbox{spaxel}}\times10^{-3}$\hspace{0.3cm} \\
 & (\ang) & & & & (\#) & \\
\hline
H$\gamma$ & 4340.5 & 1.157 &Ê$81.8\pm 9.8$  &   $130.0\pm 15.6$ & 1213 & $ 107.1\pm 12.9$ \\
He \scriptsize{II} & 4685.7 & 1.043 & $48.1\pm  3.1$  &   $73.8\pm  4.8$ & 1213 & $ 60.8\pm 4.0$ \\
H$\beta$ & 4861.3 & 1.000 & $257.0\pm 3.7$  &   $387.6\pm 5.6$ & 1213 & $ 319.5\pm 4.7$ \\
\Othree & 4958.9 & 0.976 & $224.5\pm  6.8$  &   $335.2\pm 10.2$ & 1213 & $ 276.3\pm 8.4$ \\
\Othree & 5006.8 & 0.964 & $652.6\pm 8.0$  &   $969.8\pm 11.9$ & 1213 & $ 799.5\pm 9.8$ \\
\Oone & 6300.3 & 0.718 & $152.1\pm  13.4$  &   $204.3\pm  18.1$ & 1233 & $ 165.4\pm 11.4$ \\
\Ntwo & 6548.1 & 0.682 & $60.9\pm  10.6$  &   $80.6\pm  14.0$ & 1233 & $ 65.4\pm 11.4$ \\
H$\alpha$ & 6562.8 & 0.680 & $609.2\pm 10.6$  &   $968.2\pm 14.0$ & 1233 & $ 785.3\pm 11.3$ \\
\Ntwo & 6583.5 & 0.677 & $111.4\pm  10.6$  &   $147.1\pm  14.0$ & 1233 & $ 119.3\pm 11.3$ \\
\Stwo & 6716.4 & 0.658 & $287.2\pm  9.9$  &   $376.4\pm 13.0$ & 1233 & $ 305.3\pm 10.5$ \\
\Stwo & 6732.8& 0.656 & $194.4\pm 9.8$  &   $254.6\pm 12.9$ & 1233 & $ 206.5\pm 10.4$ \\
\hline\hline
\end{tabular}
\tablefoot{Absolute observed reddened ($F(\lambda)$) and dereddened fluxes ($I(\lambda)$) are given with their corresponding errors, in units of $10^{-16}$\flux. The adopted reddening curve, $f(\lambda)$ and the average flux per spaxel $I(\lambda)_{\mbox{spaxel}}$ are also given (0.67\arcsec). The value of $c$(\Hb) is 0.18.}
\end{center}
\end{table*}

\begin{table*}[ht]
\caption[Possible emission line Fluxes of the total FoV]{\label{Tab:faint}Possible emission lines measured from the summed spectrum of the CAL 83 nebula.}
\begin{center}
\begin{tabular}{lccccc} \hline\hline\rule[0mm]{0mm}{3mm}
 \hspace{0.3cm}$\lambda_{obs}$ \hspace{0.3cm} & $f(\lambda$) & \hspace{0.3cm}$F(\lambda)$\hspace{0.3cm} & \hspace{0.3cm}$I(\lambda)$\hspace{0.3cm}  & no. spaxels & \hspace{0.3cm}$I(\lambda)_{\mbox{spaxel}}\times10^{-3}$\hspace{0.3cm} \\
 \hspace{0.4cm}(\ang) & & & & (\#) & \\
\hline
4400--4454$^*$  & 1.106 &  457.5             & 720.6 & 1213 & 594.1 \\
4659.4$^*$    & 1.050 &  $61.9\pm 17.4$  & $95.3\pm26.8$ & 1213 & $78.6\pm22.1$ \\
4663.9$^*$    & 1.048 &  $14.8\pm 15.2$  & $22.8\pm23.4$ & 1213 & $18.8\pm19.3$ \\
6462.6   & 0.694 &  $70.6\pm 10.6$    & $93.9\pm14.0$ & 1233 & $76.2\pm11.4$ \\
6495.3   & 0.689 &  $55.5\pm 10.6$    & $73.7\pm14.1$ & 1233 & $59.7\pm11.4$ \\
6532.7   & 0.684 &  $38.5\pm 10.6$    & $51.0\pm14.0$ & 1233 & $41.4\pm11.3$ \\
6634.7   & 0.670 &  $61.7\pm 10.6$    & $81.3\pm13.9$ & 1233 & $65.9\pm11.3$ \\
6683.8   & 0.663 &  $29.1\pm   8.2$    & $38.3\pm10.7$ & 1233 & $31.0\pm8.7$ \\
\hline\hline
\end{tabular}
\tablefoot{Absolute observed reddened ($F(\lambda)$) and dereddened fluxes ($I(\lambda)$) are given with their corresponding errors, in units of $10^{-16}$\flux. The adopted redding curve, $f(\lambda)$ and the average flux per spaxel $I(\lambda)_{\mbox{spaxel}}$ are also given (0.67\arcsec). Lines marked with $*$ are blends. The value of $c$(\Hb) is 0.18.}
\end{center}
\end{table*}

\subsection{Emission line morphology}

\begin{figure*}[htb]
\begin{center}
\includegraphics[width=0.7\columnwidth]{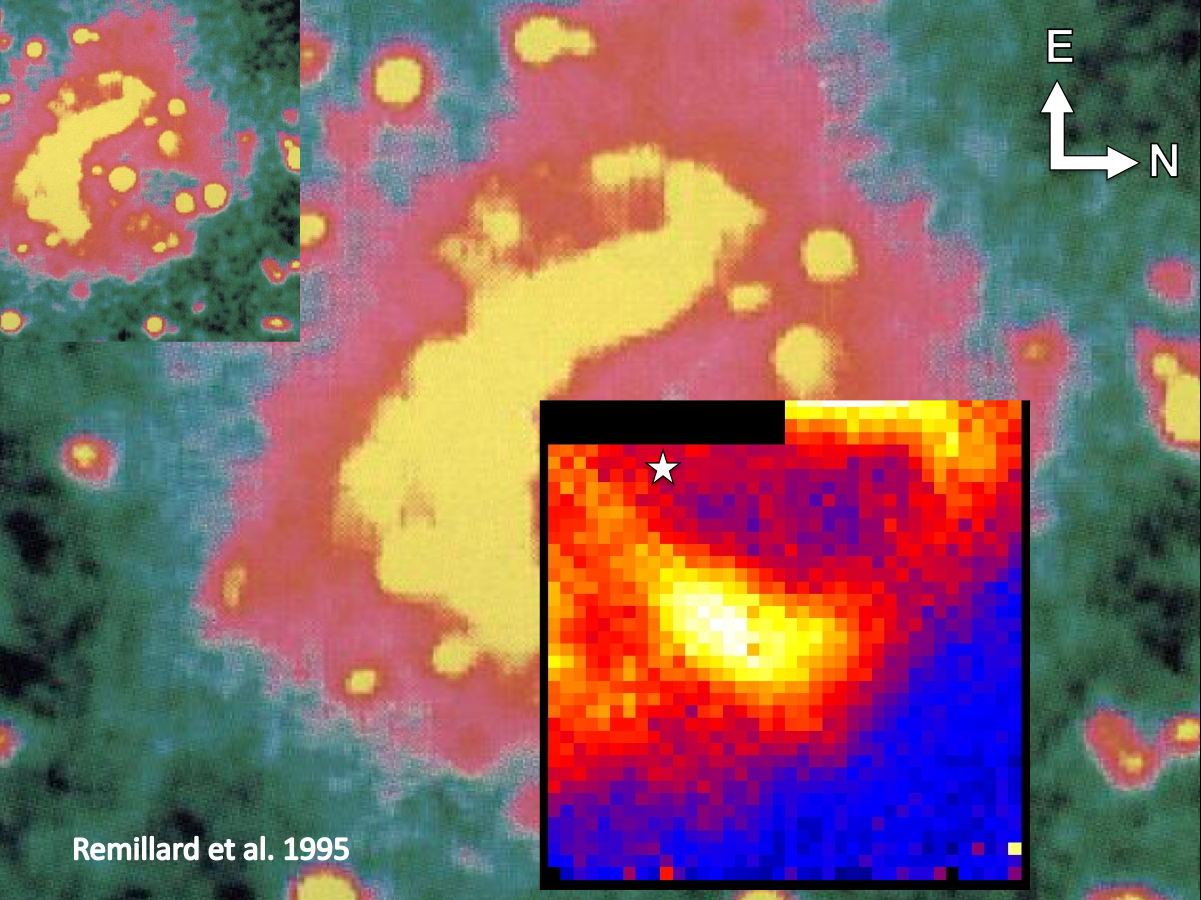}
\includegraphics[width=0.7\columnwidth]{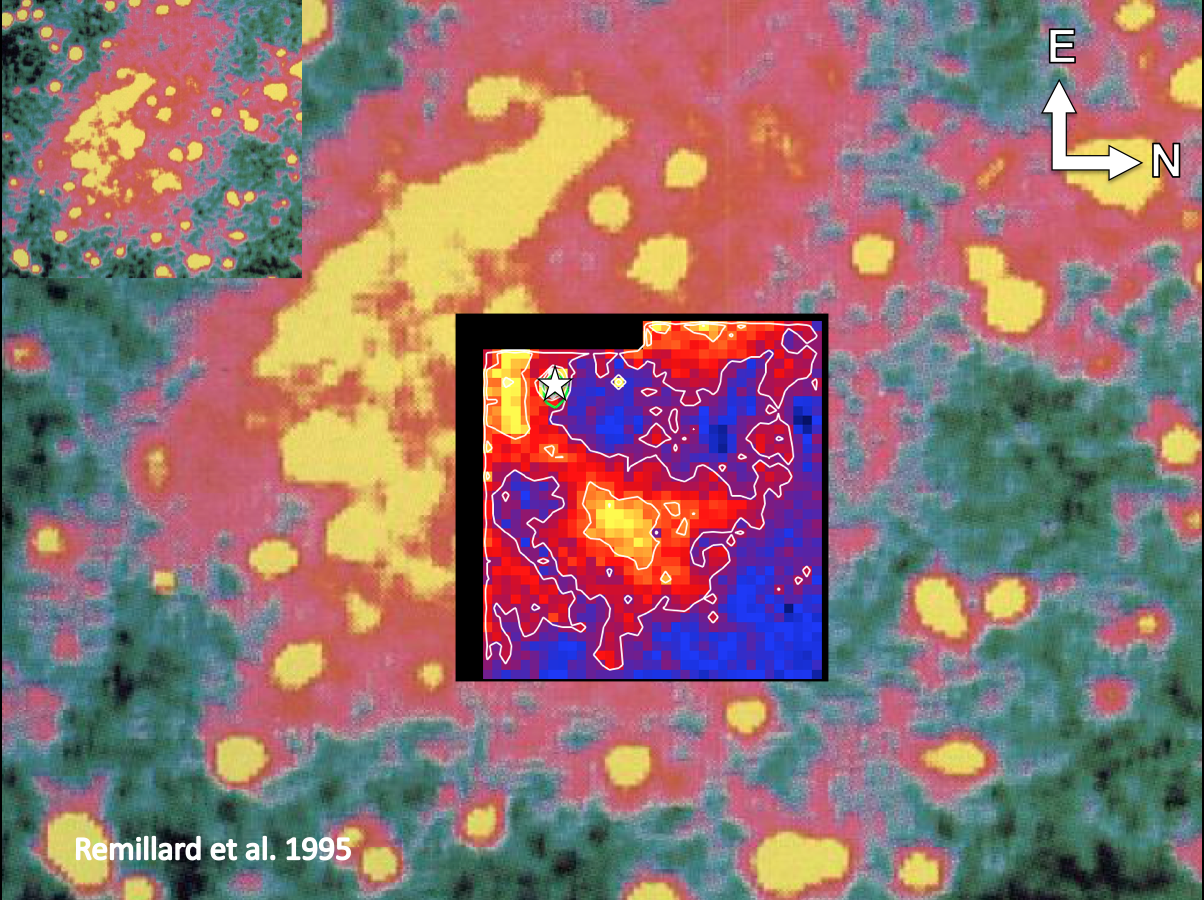}
\caption{CAL\,83 as imaged by RRM95 in \Othree\ (left) and H$\alpha$ (right). Overlaid on these are the maps made from our data. The \Othree\ image is 20\,pc across and 15\,pc in height while the \Ha\ image has somewhat larger dimensions: 25\,pc across and 18\,pc in height. CAL\,83 itself is indicated by the central white star in the images. }\label{Fig:compare}
\end{center}
\end{figure*}

\begin{figure*}[htb]
\begin{center}
\includegraphics[width=0.65\columnwidth]{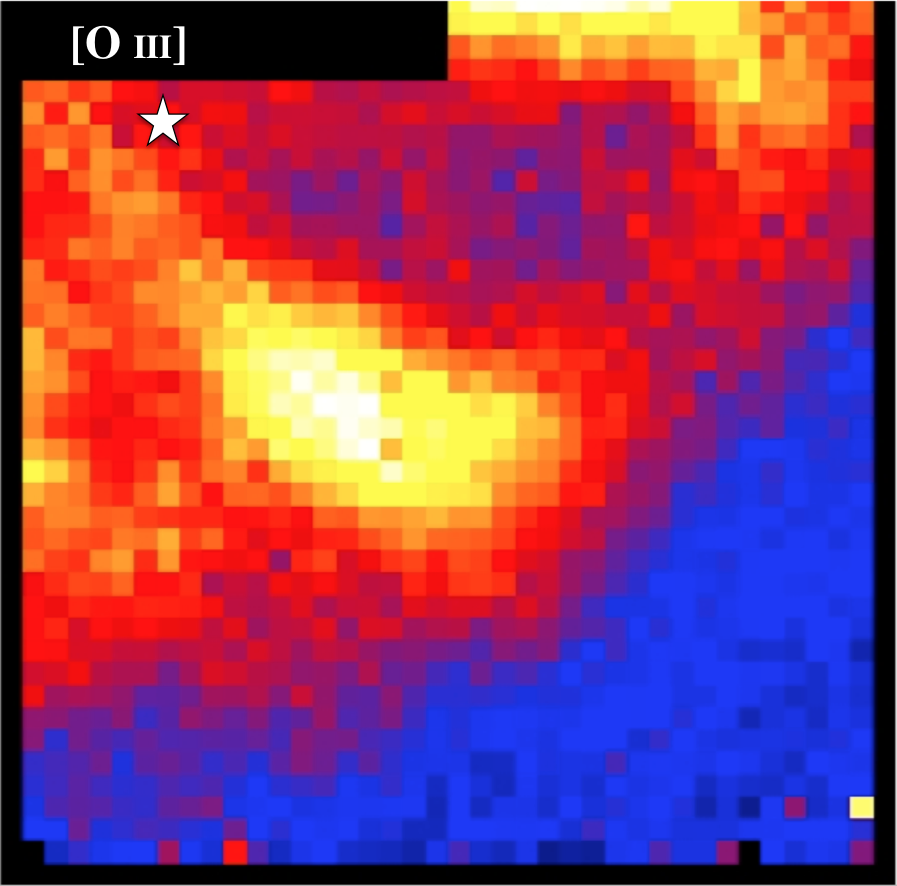}
\includegraphics[width=0.65\columnwidth]{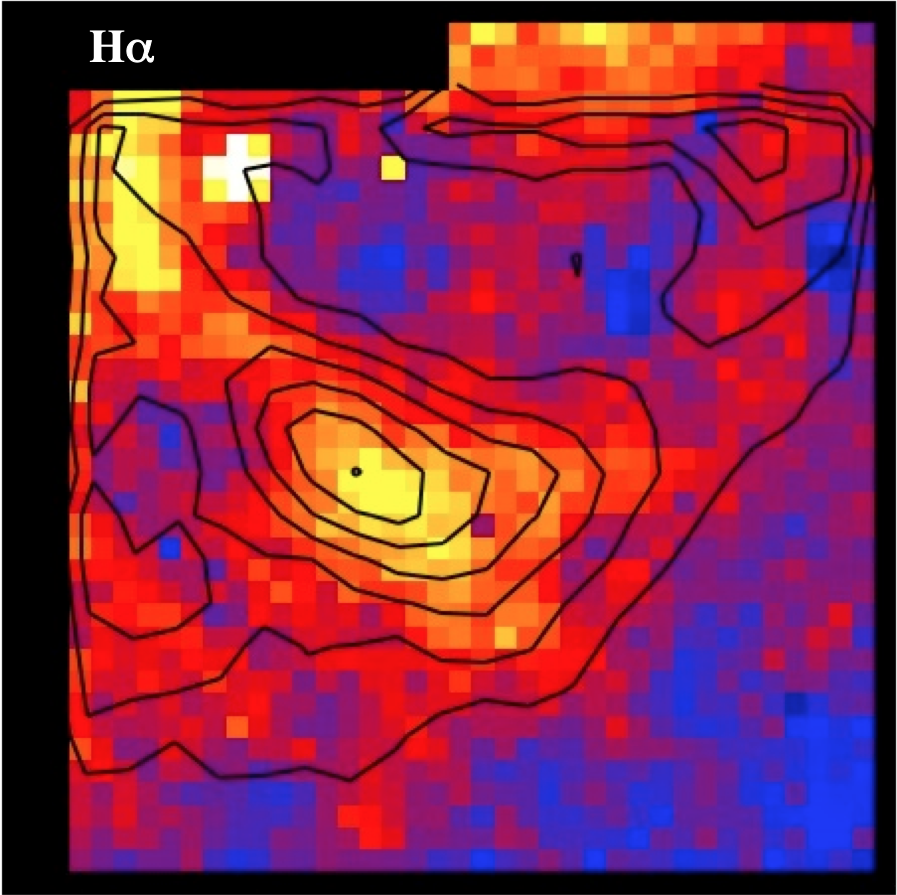}\\
\includegraphics[width=0.65\columnwidth]{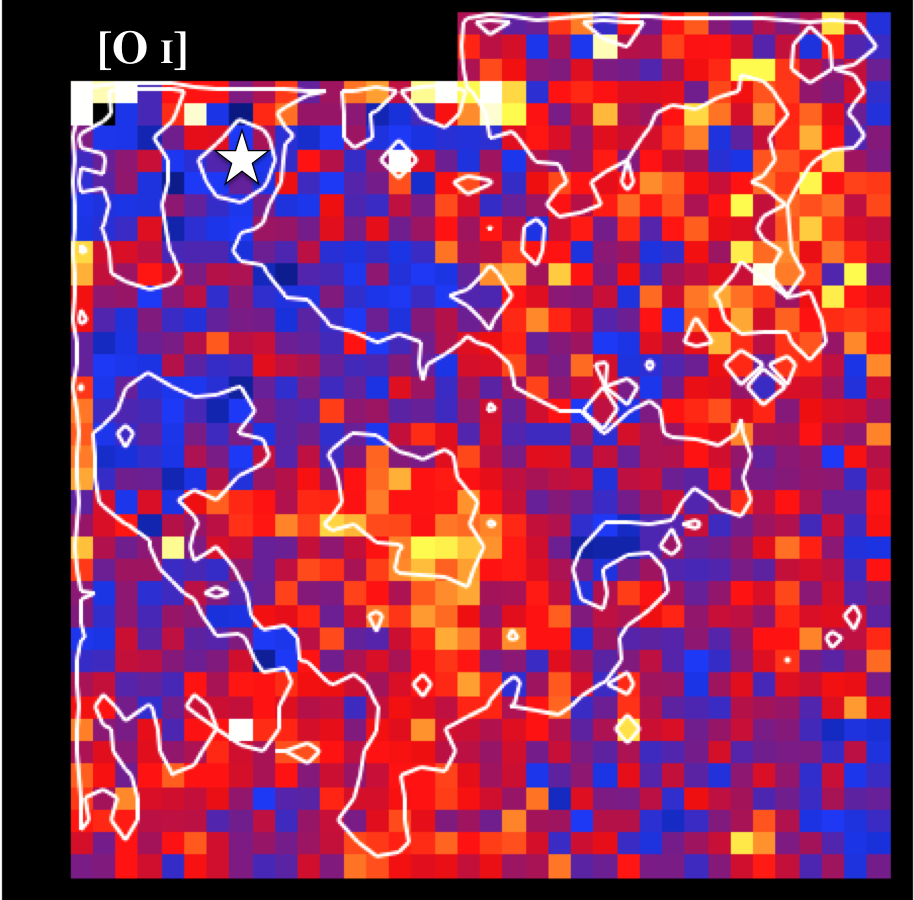}
\includegraphics[width=0.65\columnwidth]{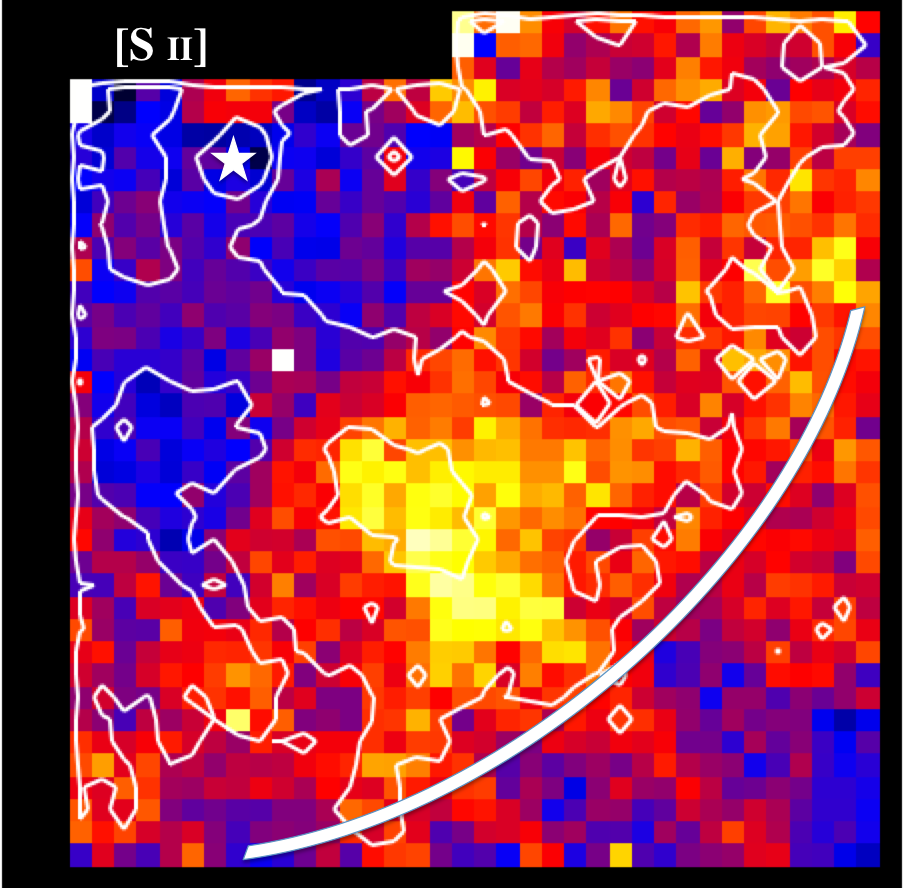}\\
\includegraphics[width=1.2\columnwidth]{ColorBar}
\caption{\textit{ Top}: [O\,{\sc iii}] map (left) and H$\alpha$ map with overplotted [O\,{\sc iii}] contours (right). The colour ranges are 0.17--0.98 for H$\alpha$ and 0.19--1.85 for [O\,{\sc iii}], in units of $10^{-16}$\flux. \textit{ Bottom}: 
[O\,{\sc i}]\,$\lambda$6300 (left), and [S\,{\sc ii}]\,$\lambda$6716 (right) maps, all with H$\alpha$ contours overplotted. In the \Stwo\ image we have indicated the transition region over which the \Stwo \ over \Ntwo\ ratio increases from 2.3 to 3.5 see section\,\ref{Sec:SIIvsNII}. The colour ranges are 0.015--0.288 for [O\,{\sc i}] and 0.106--0.454 for [S\,{\sc ii}], in units of $10^{-16}$\flux. Colour bar at the bottom indicates the normalised colour range. {\sl E is top and N is right, and the FoV is 25.5\arcsec $\times$ 25.5\arcsec\ ($7.5\times7.5$ pc$^2$). The spatial PSF is about 3 spaxels in height and 2 spaxels width. }}\label{Fig:maps}
\end{center} 
\end{figure*}

From the PAN fit results emission line maps were made. These are shown in Fig.\,\ref{Fig:maps}. The two brightest lines are included -- [O\,{\sc iii}] and H$\alpha$ (the H$\beta$ map looks much as the H$\alpha$) -- and in Fig.\,\ref{Fig:compare} our maps are compared to the images of RRM95. We see a good correspondence between our and RRM95's observations; and see that we observed about 1/4 of the full nebula. The lower panels in Fig.\ref{Fig:maps} show the ions of lower ionisation stage -- \OoneA, and \StwoA\ (\NtwoB\ shows similar behaviour as \StwoA, and hence is omitted). 

RRM95 discussed the morphology of the CAL\,83 nebula as seen in their images. In both ions it is a bright, globally symmetric but irregular nebula that extends outwards from CAL\,83 the stellar source. In \Othree\ the inner region is brightest in an arc which reaches a peak intensity near a radius of about 3.3 pc and extends to a radial distance of 7.5 pc from CAL\,83. There is also fainter emission visible out to $\sim$25\,pc from CAL\,83. The arc can also be seen in \Ha, but it is less pronounced. The region inside the arc at $<1$ pc from CAL\,83 is a local minimum in \Othree\ and less so in \Ha. Our FoV includes the inner and middle nebular regions, which are yellow and red on the RRM95 images. RRM95 find a ratio of 5:3 for the yellow:red for both mapped ions, which is close to what we find. On our maps CAL\,83 is in the SE corner.
  

The position of CAL\,83, indicated by a white star on the maps, was derived by comparing a continuum map of our FoV with the \Ha\ image of RRM95, who identified CAL\,83 as the star in the centre of the nebula. Except for H\,{\sc i}, none of the ions shown in Fig.\,\ref{Fig:maps} peaks at or around the position of CAL\,83 the stellar source. This peak in H\,{\sc i} at CAL\,83 is due to emission from the star itself (as we show later). For the \Ha\ map, uniquely among all ions, one can see that at the very SE corner of our FoV (left of the star) the emission is brightening (and we show later that it is here that the nebular He\,{\sc ii} emission can be found). It is unfortunate that our FoV is cut off here. Otherwise the morphology in \Ha\ and [O\,{\sc iii}] is similar, as can be seen by comparing the colour image to the overplotted contours: a bright peak in the centre of our FoV and secondary peaks along the E and S edges. The peak in [O\,{\sc iii}] is located closer to CAL\,83 -- the ionising source -- than that in \Ha. 

The other ions shown -- [O\,{\sc i}] and [S\,{\sc ii}] -- also peak in the same general area (the centre of our FoV) but shifted slightly further away from CAL\,83 than \Ha, and they are faint in the radial region around CAL 83 itself (i.e., in the SE quadrant). In [S\,{\sc ii}] the morphology almost appears to be that of (part of) a ring of emission with CAL\,83 at the centre.


The order of the flux peaks in our nebula, with respect to the distance from CAL\,83 [the ionising source], is: \HetwoX, \Othree, \Ha,  and then \Ntwo, \Stwo\ (and \Oone) peaking farthest out. This is in the order of their ionisation potentials, being 54.42, 35.12, 13.60 and 14.54/10.36, respectively. Looking at the overall morphology in our 1/4 field of view of the nebula, we see that the \Ntwo, \Oone\ and \Stwo\ zones are essentially coincident with each other (with \Stwo\ being the brightest), and while they also overlap with the regions of the brightest \Ha\ and \Othree\ emission, they are more-or-less absent in the immediate area around CAL\,83. So far this nebula has the appearance of a more-or-less standard ionisation nebula with a single ionising source of high temperature.


Note: the slight shift in FoV for the ions from the blue and the orange maps is due to the slight offset in the pointing for these observations. Noisy edge spaxels have been blanked out.

\subsection{He\,{\sc ii} nebular morphology}

The most interesting thing about our observations is the detection of \Hetwo\ in the nebula of CAL\,83, as it has never been observed before. The \Hetwo\ emission line has been detected in the stellar spectrum (see Sec.\,\ref{sec:heii}) and it has been predicted that it should be found in the nebulae of SXSs (e.g., \citealt{Chiang1994}). The stellar line is discussed in the next section, here we discuss the He\,{\sc ii} morphology.

\begin{figure}
\begin{center}
\includegraphics[width=0.9\columnwidth]{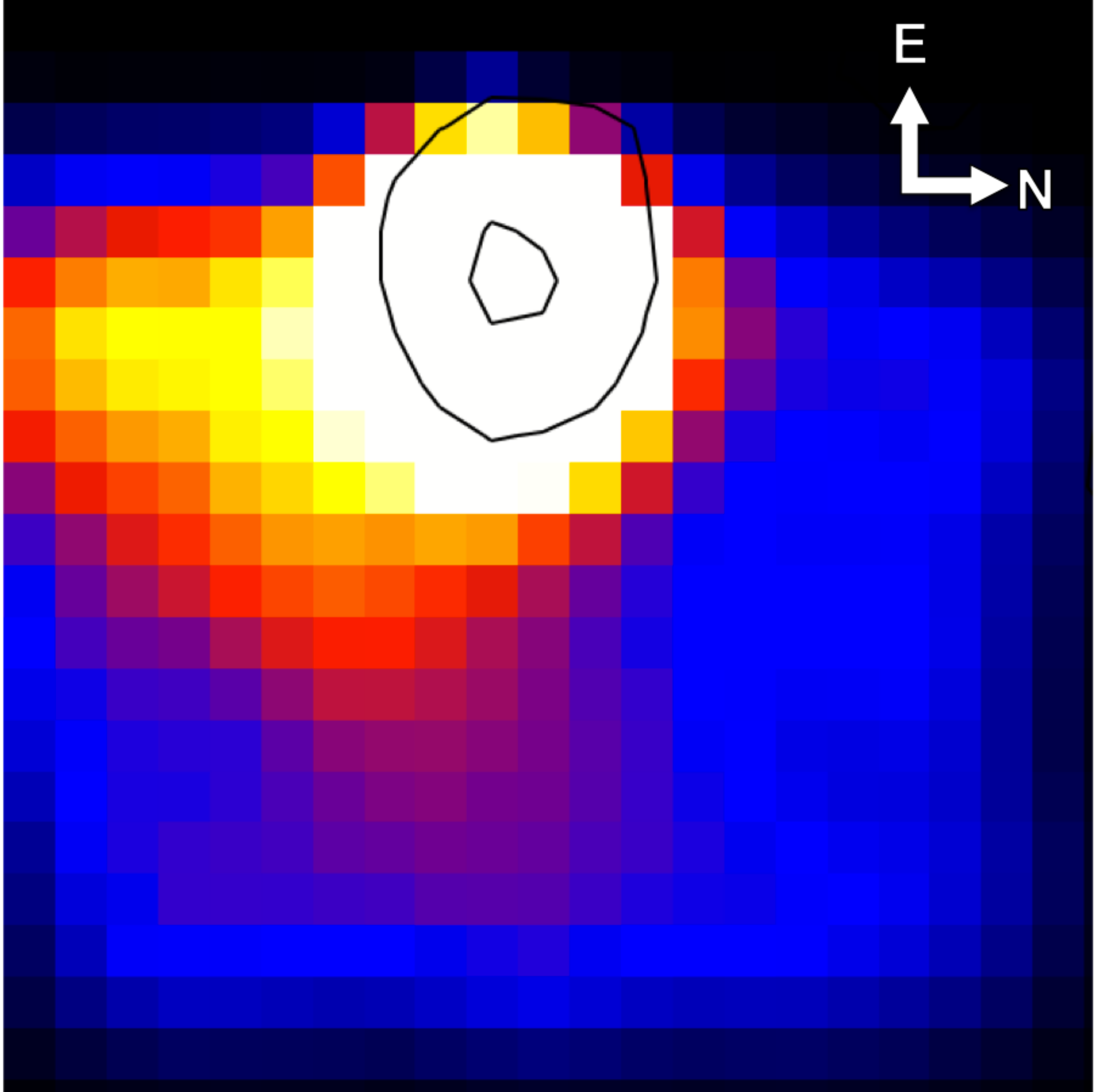}
\includegraphics[width=0.9\columnwidth]{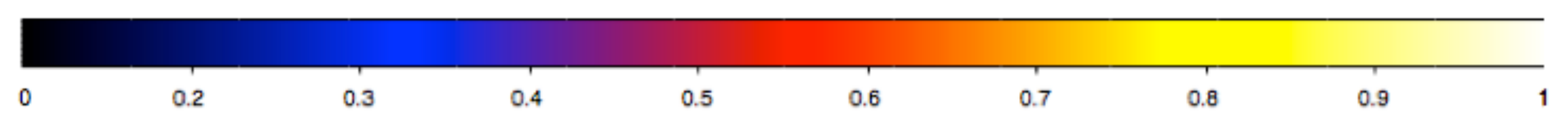}
\caption{ \Hetwo\ flux map, Gaussian-smoothed over a $3\times3$ pixel box. Overlaid in black are the contours representing the FHWM and FWZM [deduced from the CAL 83 stellar continuum emission]. Note that we show here {\sl only one quadrant} of our FoV, that where \HetwoX\ emission was detected. The colour ranges from 0.003 to 0.49 $\times10^{-16}$\flux. The colour bar represents the entire normalised colour range. }
\label{Fig:Helium_flux}
\end{center}
\end{figure}

The map of He\,{\sc ii} is presented in Fig. \ref{Fig:Helium_flux}. We could only measure this emission line in each spaxel for one quadrant, the one containing CAL\,83 itself. On the map one can see a bright region, which is where CAL\,83 is located. To confirm this, and to check whether this bright emission arises from a point source (i.e., CAL\,83 itself) we compared the spatial distribution of He\,{\sc ii} at the location of CAL\,83 to the spatial distribution of CAL\,83 from a continuum spectral region (i.e., from the star itself). These continuum contours are shown also in Fig. \ref{Fig:Helium_flux}. It is clear that the brightest emission in He\,{\sc ii} arises from a point source.

There is faint, but obvious He\,{\sc ii} emission to the immediate south of CAL\,83 (where H$\alpha$ also shows a peak; see Fig.\,\ref{Fig:maps}). Although we have not, unfortunately, mapped the entire surroundings of CAL\,83, it is clear that this He\,{\sc ii} emission is not symmetrically distributed about CAL\,83 and is fairly confined. With an instrumental broadening of $\sim100$ \kms, the nebular line is spectrally unresolved and shows no radial velocity gradient that we can measure. 
The flux of He\,{\sc ii} coming exclusively from the bright \HetwoX\ nebula (see L-shape in Fig.\,\ref{Fig:ratio}), excluding CAL\,83 itself, is  $(29.98\pm1.4)\times10^{-16}$\,\flux. Even though this is the brightest zone in the \HetwoX\ nebula,  this is not all the \Hetwo\ flux in the FoV--the total flux measured over our FoV is $(73.8\pm4.8)\times10^{-16}$\,\flux--but unfortunately we could not measure the line per spaxel outside of the area shown in Fig.\,\ref{Fig:Helium_flux}. 


We note that we did not detect any \Heone\ in our FoV.

\subsection{He\,{\sc ii} (and H$\alpha$) stellar line}
\label{sec:heii}
\begin{figure*}
\begin{center}
\includegraphics[width=1\columnwidth]{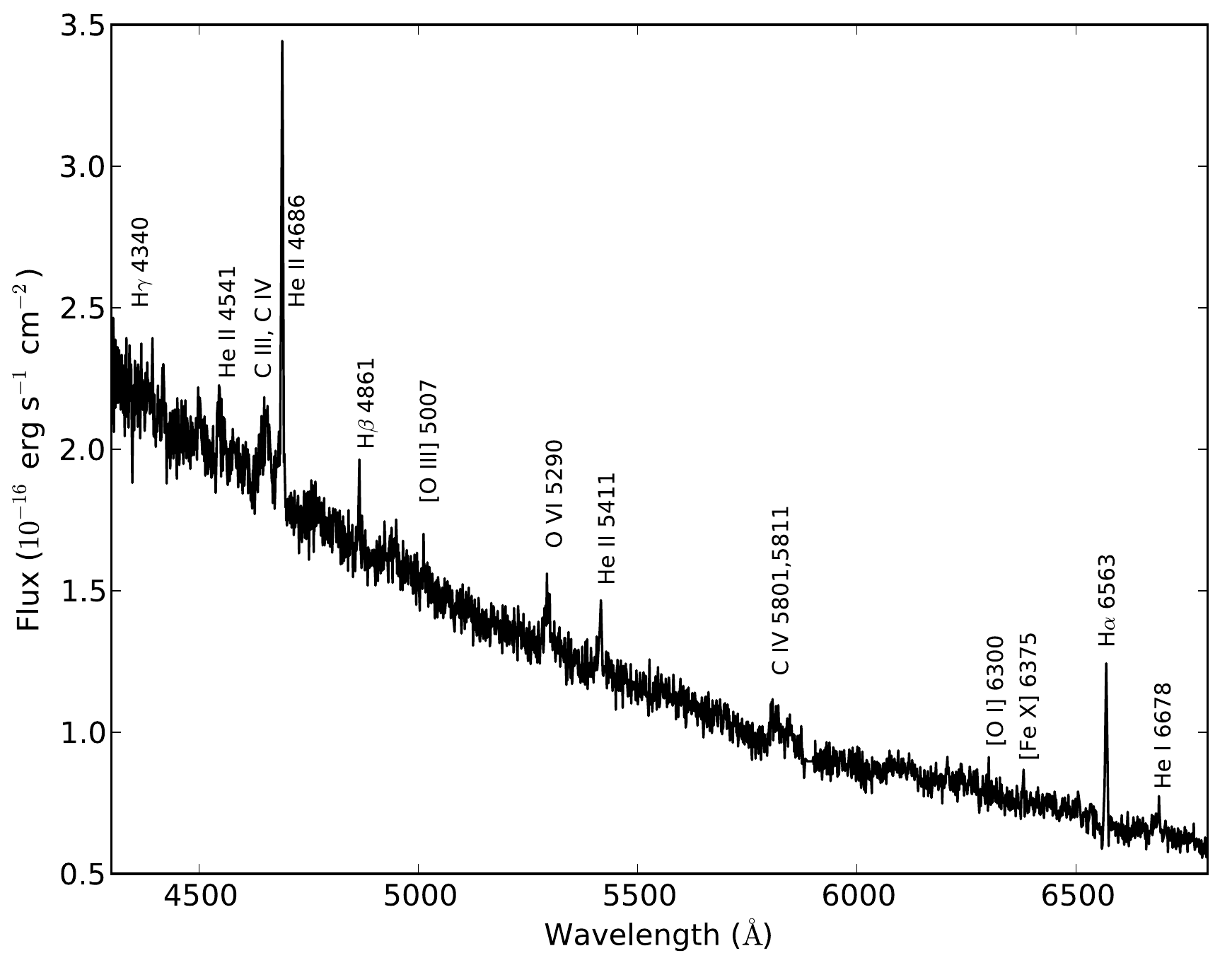}
\includegraphics[width=1\columnwidth]{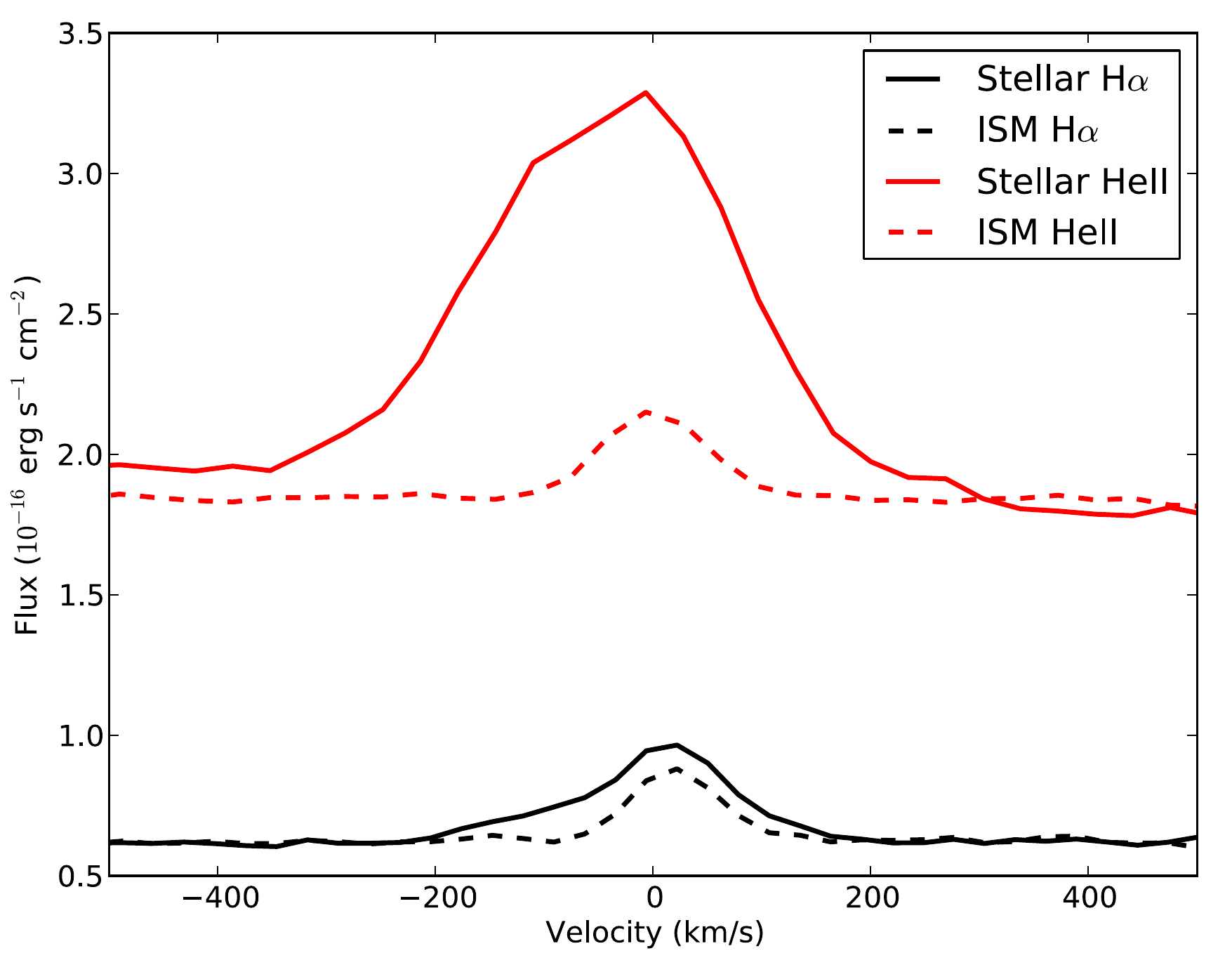}
\caption{\textit{ Left:} CAL\,83 {\sl stellar} spectrum extracted from 6 spaxels that include the source. The ISM contribution has been removed and the detected emission lines are labelled. \textit{ Right:} a zoom on the stellar He\,{\sc ii}\ and \Ha\ line. The spectrum is plotted in velocity space, normalised to the velocity of the LMC: 270\,km/s. Shown for comparison are the ISM lines, normalised to the same area over which the stellar source extends, and taken from the area where He\,{\sc ii} emission can actually be found (see Fig.\,\ref{Fig:Helium_flux}). Note that the ISM emission has not been subtracted from the spectrum. {\sl The spectra are not dereddened}.}
\label{Fig:Star}
\end{center}
\end{figure*}

We have extracted the stellar spectrum of CAL\,83 from the six spaxels that contain the flux from the central star, CAL\,83. The ISM contribution has been removed from this full-wavelength range spectrum, shown on the left of Fig.\,\ref{Fig:Star}. In the right of Fig.\,\ref{Fig:Star} we show a zoom-in on He\,{\sc ii} and \Ha, which are the only two emission lines in our spectrum that are also bright peaks at the position of the star. For this zoom, the ISM continuum has {\em not} been removed: instead we show for comparison the area-normalised spectrum of the ISM in these two emission lines. The stellar line is much broader than the nebular line and has more emission in its blue wing. While we cannot unambiguously establish whether there is any nebular emission at the location of the stellar source, it is clear that any such contamination is small for He\,{\sc ii} but more important for \Ha. Our measured fluxes are not corrected for any ISM contamination, as we cannot be sure that ISM emission is blended with stellar emission. 

The stellar spectrum in the left panel of Fig. \ref{Fig:Star} is similar to the spectra published by \citet{Crampton1987} and \citet{Cowley1998}. On the blue side of He\,{\sc ii} we find a blend of carbon lines (C\,{\sc iii}, C\,{\sc iv}), these being first noted by \citet{Pakull1985}. This emission, together with O\,{\sc vi} at 5290\,\AA, implies a very high degree of ionisation (IP$_{\mbox{O\,\scriptsize{VI}}} = 114 $ eV). We also note the detection of the $\left[\mbox{Fe\,\scriptsize{X}}\right] \lambda6375$ coronal line in emission as first observed by \citet{Oliveira2006}. \\

Noting the variable position and strength of He\,{\sc ii}, \citet{Crampton1987} argued that He\,{\sc ii} and the C and O emission features arise from the accretion disc of the CAL\,83 system. This was later confirmed by \citet{Cowley1998}, who finds evidence that the emission is linked to matter being ejected in bipolar flows, which may arise from a 69-day precessing accretion disk. For our observations we find FWHM $\sim$$305$\,\kms\ agreeing well within the range of values found by \citet{Smale1988}, i.e., 233 to 376\,\kms\ over a time period of one year. The FWZM extends to a velocity of 1150\,\kms\ while \citet{Cowley1998} state a velocity of $\sim2450$\,\kms\ derived from a single observation.

\citet{Smale1988} deduced a time-averaged luminosity from the He\,{\sc ii}\ stellar line, $L_{*}$(\HetwoX) = $3.6\times10^{33}$ (d/55\,kpc)$^2$\,\ergs. They observe a variation in the strength of He\,{\sc ii} of a factor of 4 in equivalent width over a 1 year period. We measure a flux of $(11.3\pm0.2)\,\times10^{-16}$ \flux, and converting this to a luminosity ($L=4\pi d(\mbox{cm})^2 F$(\flux) and abopting a distance of 55\,kpc) we obtain $L_{*}$(\HetwoX)$=(4.1\pm0.1)\,\times10^{32}$ \ergs. Although this is a (slight) lower limit to the actual stellar flux, our FoV just clips the edge of the star, it is about a factor of 10 less than the time averaged value.  (Remember that the measured flux contains contributions from the star as well as from any ISM located between CAL\,83 and us.) 

We also find an extended blue wing in \Ha\ (FWHM\,$\sim270$\,\kms). We measure a flux of $(6.5\pm0.4)\,\times10^{-16}$ \flux\ from the \Ha\ stellar line, corresponding to a luminosity of $L_{*}$(\Ha)$=(2.4\pm0.4)\,\times10^{32}$ \ergs. We could not find any published values to compare to. 

\section{Nebular diagnostics}
\label{sec:ionstage}

Using certain emission line ratios we have placed limits on the electron temperature and density of the CAL\,83 nebula. We then use various common emission line ratios to study the variation of ionisation across the nebula.

\subsection{T$_{\rm e}$ and n$_{\rm e}$}

For photoioinised nebulae one can use the ratio of the sulphur lines (\StwoA/\StwoB) to determine a density. The median flux ratio over our FoV is 1.44 with an uncertainty of $\pm{0.14}$ and a standard deviation of 0.2: so at best any variation in the ratio over the FoV is small, (and we can also see that it is random).
Following \citet{Osterbrock1989}, with an assumed electron temperature of 10,000\,K, this corresponds to an electron density of less than 10 particles per \,cm$^{3}$. The average ISM density in the inner nebula (i.e., characterised by a radius of $\sim$$7.5$ pc) was estimated by RRM95 to be in range of 4--10\,cm$^{-3}$, in agreement with our findings. We assumed an electron temperature as we do not have all the [O\,{\sc iii}] or [N\,{\sc ii}] lines which are necessary for a more self-consistent measurement, although the lack of [O\,{\sc iii}]\,$\lambda$4363 (i.e., the 1-$\sigma$ limit) leads to an upper limit T$_{\rm e} \simeq14,500$\,K. 

\subsection{Ion-Ion maps: morphology}\label{Sec:SIIvsNII}
 
One can characterize the spectral energy distribution (SED) of a source that is ionising a nebula by comparing the emission line fluxes from a modelled nebula (with input parameters of the nebular temperature and density, as well as source SED) to what is observed. However, not only must the nebular conditions be known, but one needs to be sure that all ions stages have been measured: this is not always possible, especially when only a limited wavelength coverage is available. Alternatively, a qualitative measure of the hardness of the radiation that is ionising the nebula can be obtained by comparing the emission from lines of ions the lower ionisation potentials (e.g. [O\,{\sc i}], \Ntwo\ and [S\,{\sc ii}]), to the lines of ions with higher ionisation potential, such as He\,{\sc ii} and [O\,{\sc iii}]. The diagnostic ratios we will use to do this are \OthreeB/\Hb, \NtwoB/\Ha\ and \StwoC/\Ha. The first ratio is a good reddening-free indicator of the mean level of ionisation (radiation field strength), since [O\,{\sc iii}] arises from an ion of a relatively high ionisation potential (35.1eV for O$^{+}$ $\rightarrow$ O$^{2+}$). The other two ratios are related to gas ionised at a lower energy level: the [S\,{\sc ii}] and  [N\,{\sc ii}] lines arise from ions of lower energies (S$^0$ $\rightarrow$ S$^+$=10.36eV, N$^0$ $\rightarrow$ N$^+$=14.54eV).

\begin{figure*}
\begin{center}
\includegraphics[width=0.65\columnwidth]{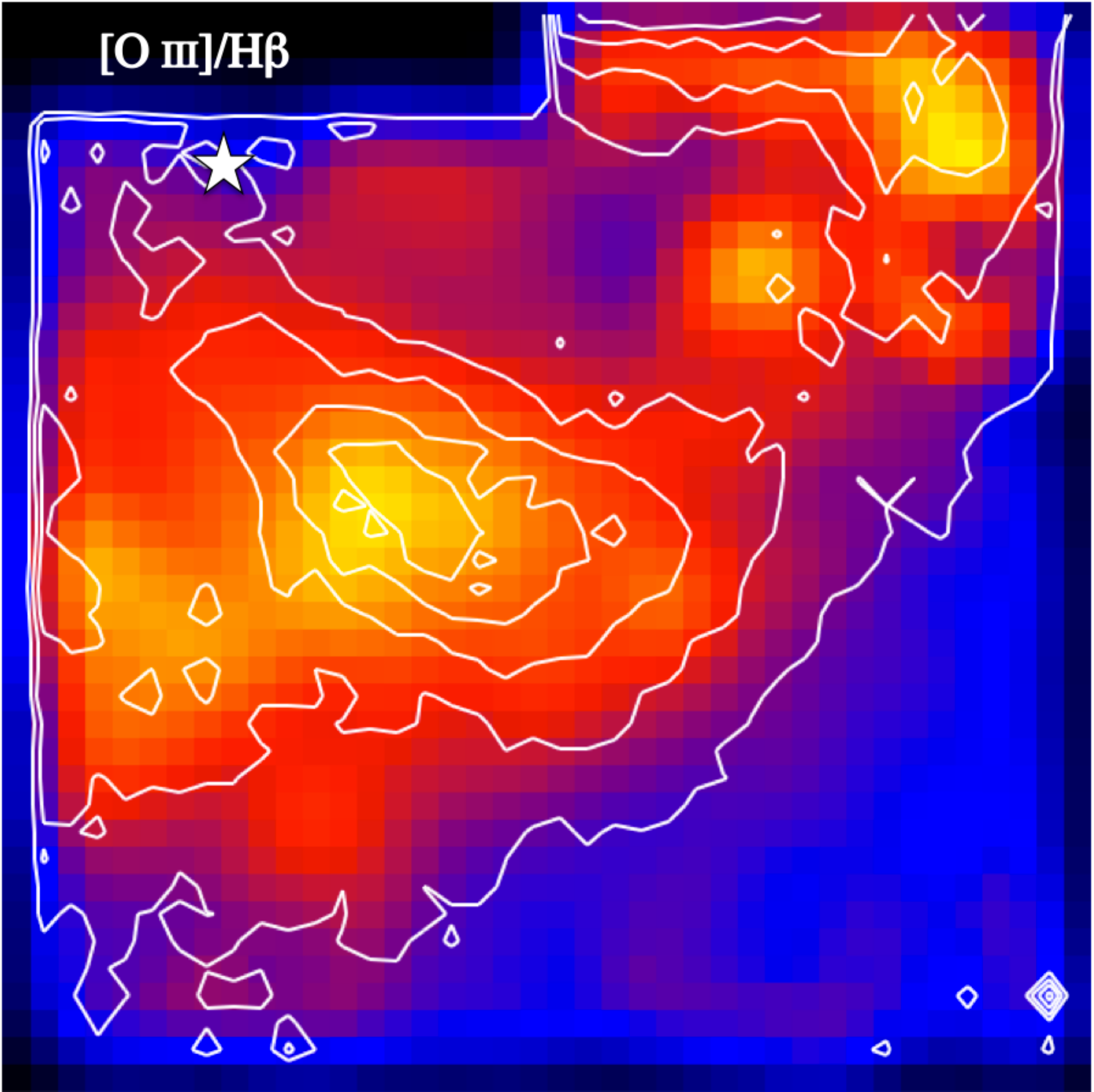}
\includegraphics[width=0.65\columnwidth]{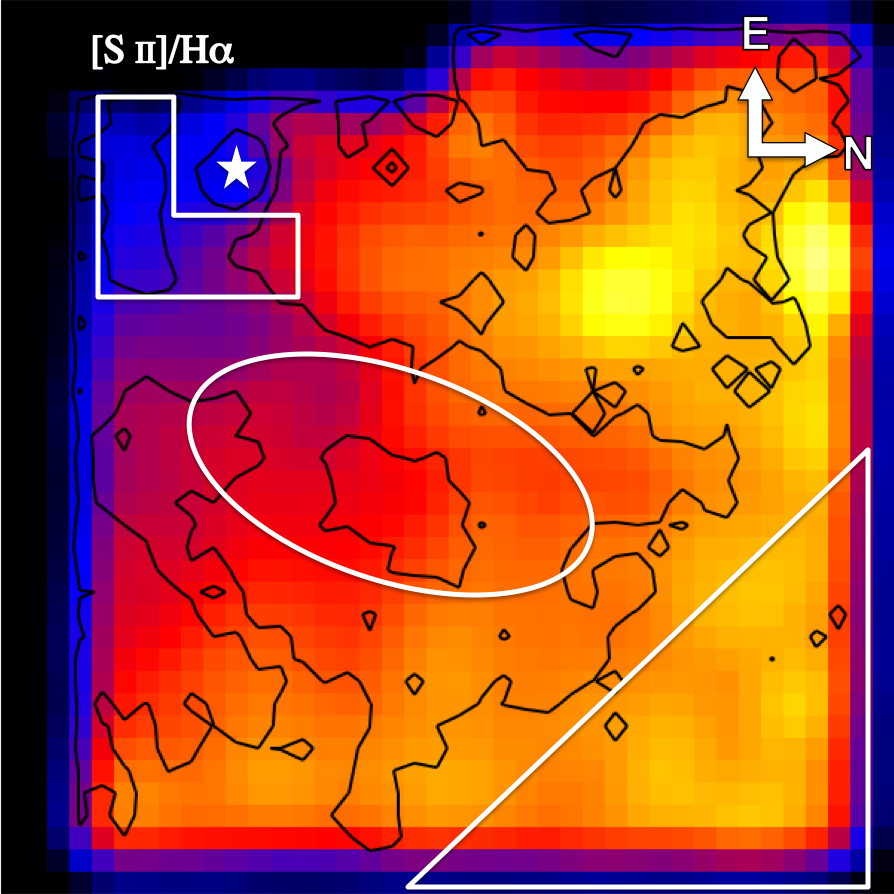}\\
\includegraphics[width=0.65\columnwidth]{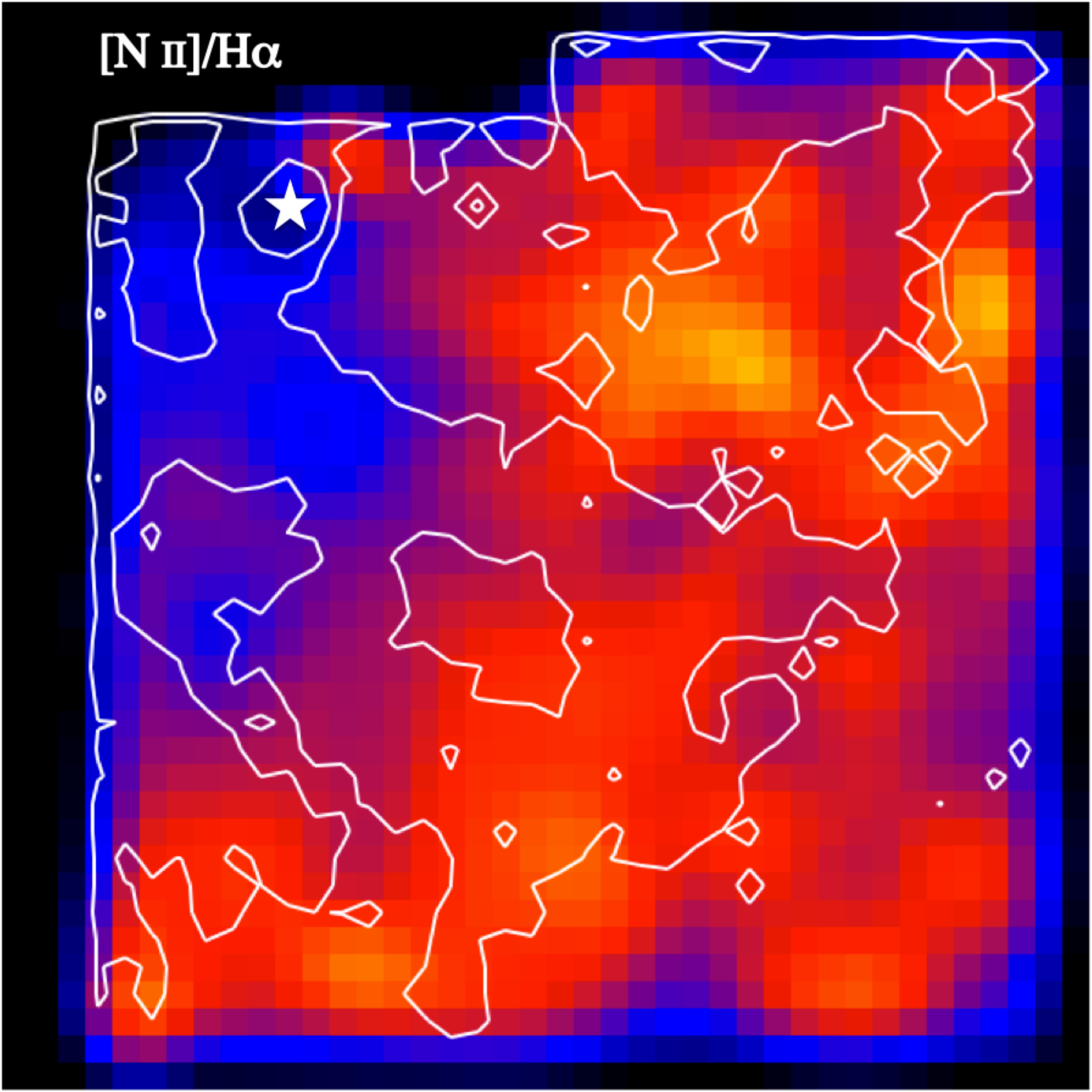}
\includegraphics[width=0.65\columnwidth]{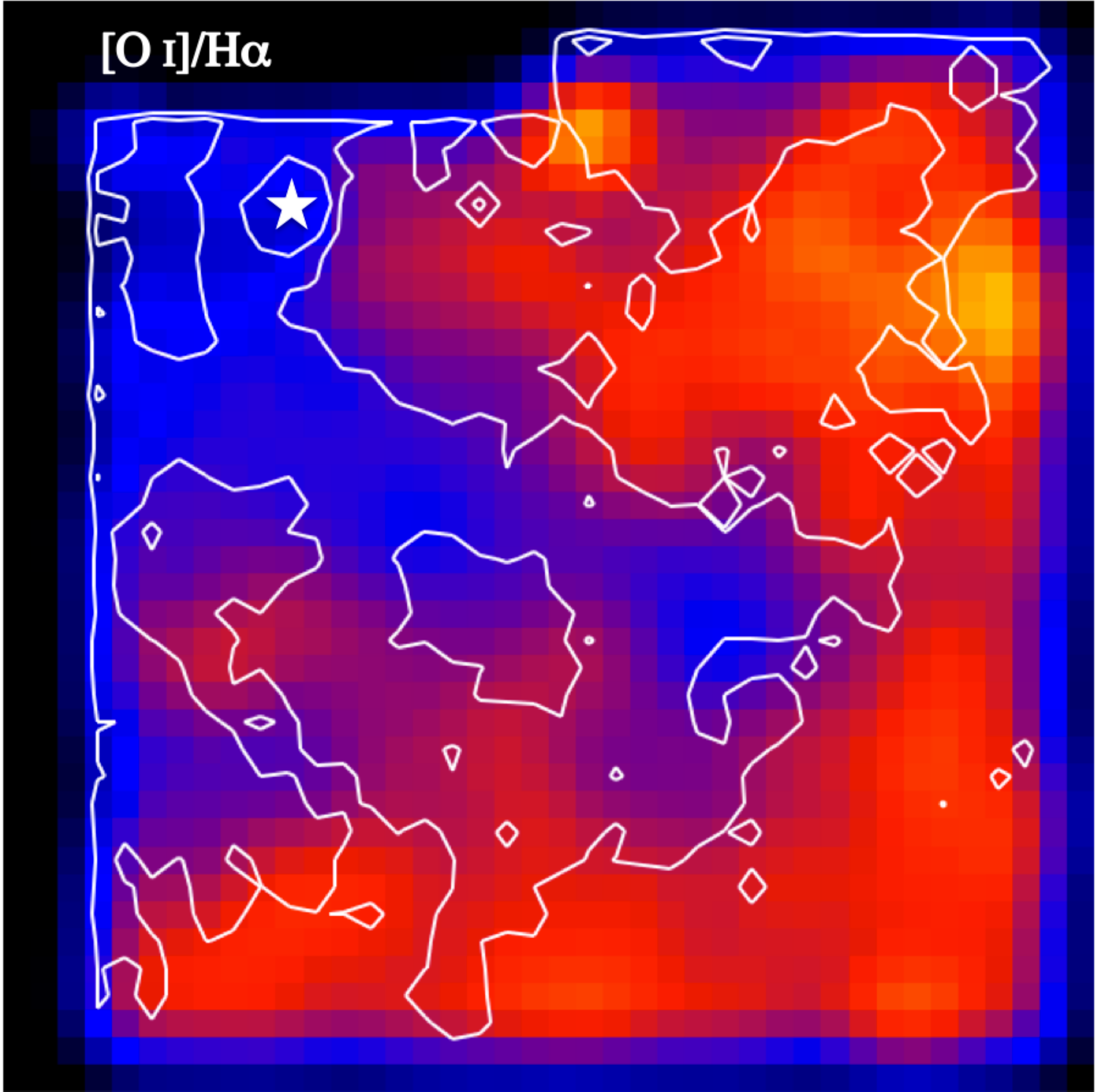}\\
\includegraphics[width=1.2\columnwidth]{ColorBar}
\caption{ \textit{ Top left}: Ratio map of \OthreeB/\Hb\ with the \OthreeB\ contours overplotted as reference and the position of CAL\,83 is noted with a white star. The colour range is 0.05--6.5. \textit{ Top right}: Ratio map of \StwoC/\Ha\ with the \Ha\ contours overplotted. The colour range is 0.15--1.2. The white shapes show the regions from which we measured line ratios: the triangle corresponds to the faint nebula region, the oval to the central region and the L-shape to the \HetwoX\ region. \textit{ Bottom left}: Ratio map of \NtwoB/\Ha\ with the \Ha\ contour overplotted. The colour range is 0.05--0.3. \textit{ Bottom right}: Ratio map of \OoneA/\Ha\ overplotted by the \Ha\ contours. The colour range is 0.03--0.47. The colour bar represents the normalised colour range. All maps are Gaussian-smoothed over a $3\times3$ pixel box. {\sl E is up and N is right and the FoV is 25.5\arcsec $\times$ 25.5\arcsec\ ($7.5\times7.5$} pc$^2${\sl )}.} 
\label{Fig:ratio}
\end{center}
\end{figure*}

Now we describe what we see in Figs\,\ref{Fig:maps} and \ref{Fig:ratio}, and relate this to the ionisation potentials of the ions. \\
\Stwo\ will rise in flux when h$\nu$ reaches 10.36 eV. Roughly speaking, one would expect [S\,{\sc ii}] to remain ionised throughout the transition region between H$^+$ and H$^0$ (IP 13.6ev), and this we do see: [S\,{\sc ii}] is found in the region where H$\alpha$ lies but extends further out, where the H$\alpha$ flux drops. \\
Nitrogen has a first ionisation potential of 14.54 eV, slightly higher than that for H$\alpha$, and so, assuming that a transition region exists between singly ionised hydrogen and neutral hydrogen (as it does for CAL\,83), the [S\,{\sc ii}] emission will become dominant over the [N\,{\sc ii}] emission as one moves through this transition region going away from the ionising source \citep{Seaton1970}. We have some evidence of this: on the map of [S\,{\sc ii}] in Fig.\,\ref{Fig:maps}, we have drawn an arc where, as one moves away from the ionising source, the mean [S\,{\sc ii}]/[N\,{\sc ii}]  ratio increases from 2.5 to 3.5. This border is coincident with the region where the H$\alpha$ emission drops. \\
To have [O\,{\sc i}]  emission one requires neutral oxygen, and photons of h$\nu>13.6$eV (IP for ionising to the first level) will decrease its abundance and hence the [O\,{\sc i}] flux. Hence the [O\,{\sc i}] flux should be strongest farthest away from the ionising source: while [O\,{\sc i}]  is faint on our maps, it is found throughout the region where \Stwo\ is, and it is likely that for [O\,{\sc i}]  (as well as [S\,{\sc ii}]  and [N\,{\sc ii}])  we have not observed the entirely of their emission regions, which will extend much further beyond the edge of our maps.\\

The top left panel in Fig.\,\ref{Fig:ratio} shows that the peak of the \Othree/\Hb\ ratio lies near the central region that is bright in all ions, and that it is fainter around CAL\,83 itself. The more highly ionised region, characterised by a high \Othree/\Hb\ ratio, fills the ``inner nebula" of RRM95 (i.e., the S--E diagonal half of our FoV). The regions of low ionisation states, characterised by relatively high \Stwo/\Ha\ and \Ntwo/\Ha\ ratios (see Fig.\,\ref{Fig:ratio}) are found over and beyond the edge heading towards the outer nebular region of RRM95 (i.e. the N--W diagonal half of our FoV). The map of \OoneA/\Ha, and also \Stwo/\Ha, shows clearly that these ratios are greatest where the \Ha\ flux is lowest: arising from the edges of the ``nebula" where the ionisation of H is dropping. In objects which are photoionised by a spectrum containing a large fraction of high-energy photons these zones in which the ionisation of H is dropping are quite extended \citep{Veilleux1987}.

\subsection{Ion-ion ratios: diagnostic diagrams}

\begin{table*}[ht]
\caption[Diagnostic ratios.]{\label{Tab:diagnostics} Diagnostic emission line ratios}
\begin{center}
\begin{tabular}{lccccccc} \hline\hline\rule[0mm]{0mm}{3mm}
\multirow{2}{*}{Region (no. spaxels)} & \multirow{2}{*}{log$\frac{\mbox{\Othree}}{\mbox{\Hb}}$} & \multirow{2}{*}{log$\frac{\mbox{\Oone}}{\mbox{\Ha}}$} &  \multirow{2}{*}{log$\frac{\mbox{\Stwo}}{\mbox{\Ha}}$} &  \multirow{2}{*}{log$\frac{\mbox{\Ntwo}}{\mbox{\Ha}}$} &  \multirow{2}{*}{log$\frac{\mbox{\HetwoX}}{\mbox{\Hb}}$} &  \HetwoX & \Hb \\
  & & & & & & {\scriptsize ($\times10^{-16}$\flux)} & {\scriptsize ($\times10^{-16}$\flux)} \\
\hline
Nebula (Full FoV: 1213/1233) & 0.39 & -0.68 & -0.19 & -0.82 & -0.72 & 73.79 & 387.6 \\
\HetwoX\ region (SE: 79) & 0.49 & -0.82 & -0.31 & -0.85 & -0.23 & 29.98 & 55.78 \\
Central region (171) & 0.57 & -0.60 & -0.14 & -0.77 & -0.70 & 11.66 & 59.43 \\
Faint region (NW: 556) & 0.31 & -0.57 & -0.12 & -0.76 & -0.89 & 18.43 & 141.3 \\
\hline\hline
\end{tabular}
\tablefoot{The diagnostic ratios for different regions in the nebula along with the measured \HetwoX\ and \Hb\ flux and the number of spaxels over which the fluxes were measured. The `nebula' values correspond to the fluxes measured from our full FoV using 1213 and 1233 spaxels for respectively the blue and orange observations. }
\end{center}
\end{table*}

We now take a more quantitative look at the line-ratio maps. Emission line-ratio values for interesting regions in the nebula can be found in Table \ref{Tab:diagnostics}. The regions are drawn on the upper right plot in Fig. \ref{Fig:ratio} (\Stwo/\Ha). The triangle-shaped region corresponds to the ``faint region", the oval to the ``central region" and the L-shaped region to the ``\HetwoX\ region".
We can compare our derived ratios with those of different astronomical sources by placing them in a \emph{diagnostic diagram}. A diagnostic diagram helps to classify the dominant energy source in emission line sources using optical spectral diagnostics in a qualitative way. The diagrams are based on empirical findings and can be compared quantitatively to theoretical expectations. Our diagnostic diagrams combine a measure for the hardness of the ionising radiation (\OthreeB/\Hb) with measures for the low ionisation degree (\OoneA/\Ha, \StwoC/\Ha\ and \NtwoB/\Ha) in the nebula. 

\begin{figure*}
\centering
\includegraphics[width=0.95\columnwidth]{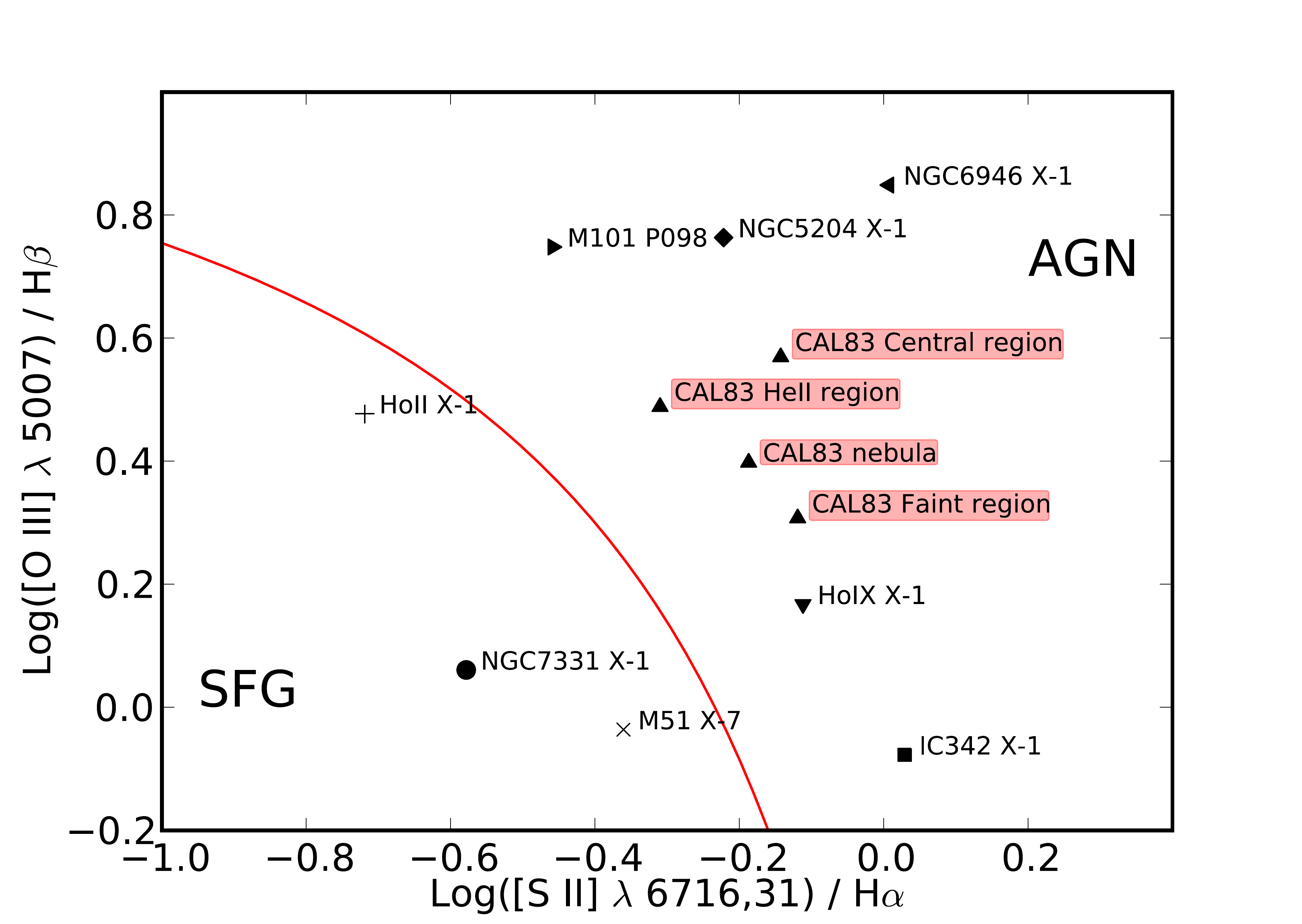}
\includegraphics[width=0.95\columnwidth]{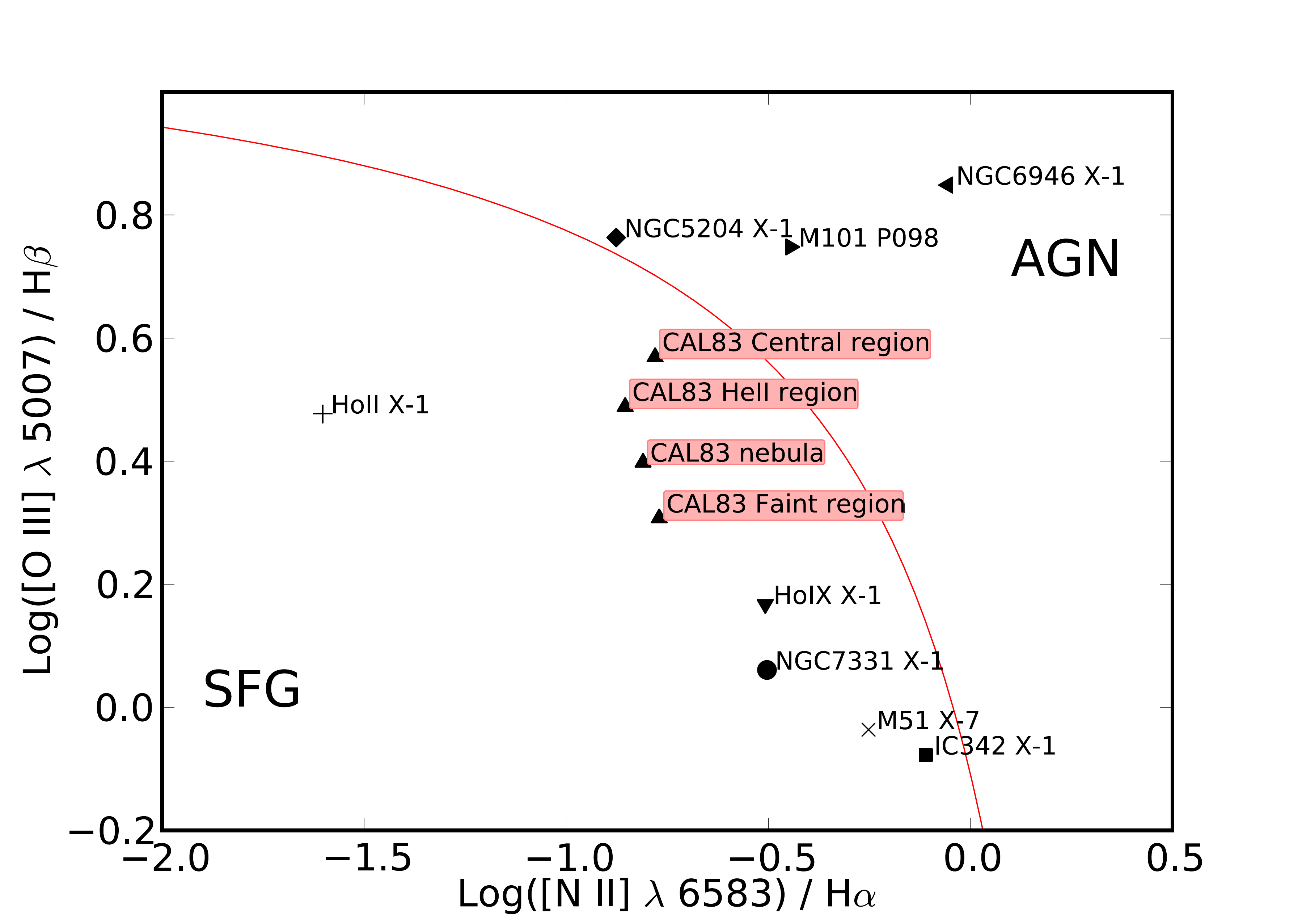} \\
\includegraphics[width=1.2\columnwidth]{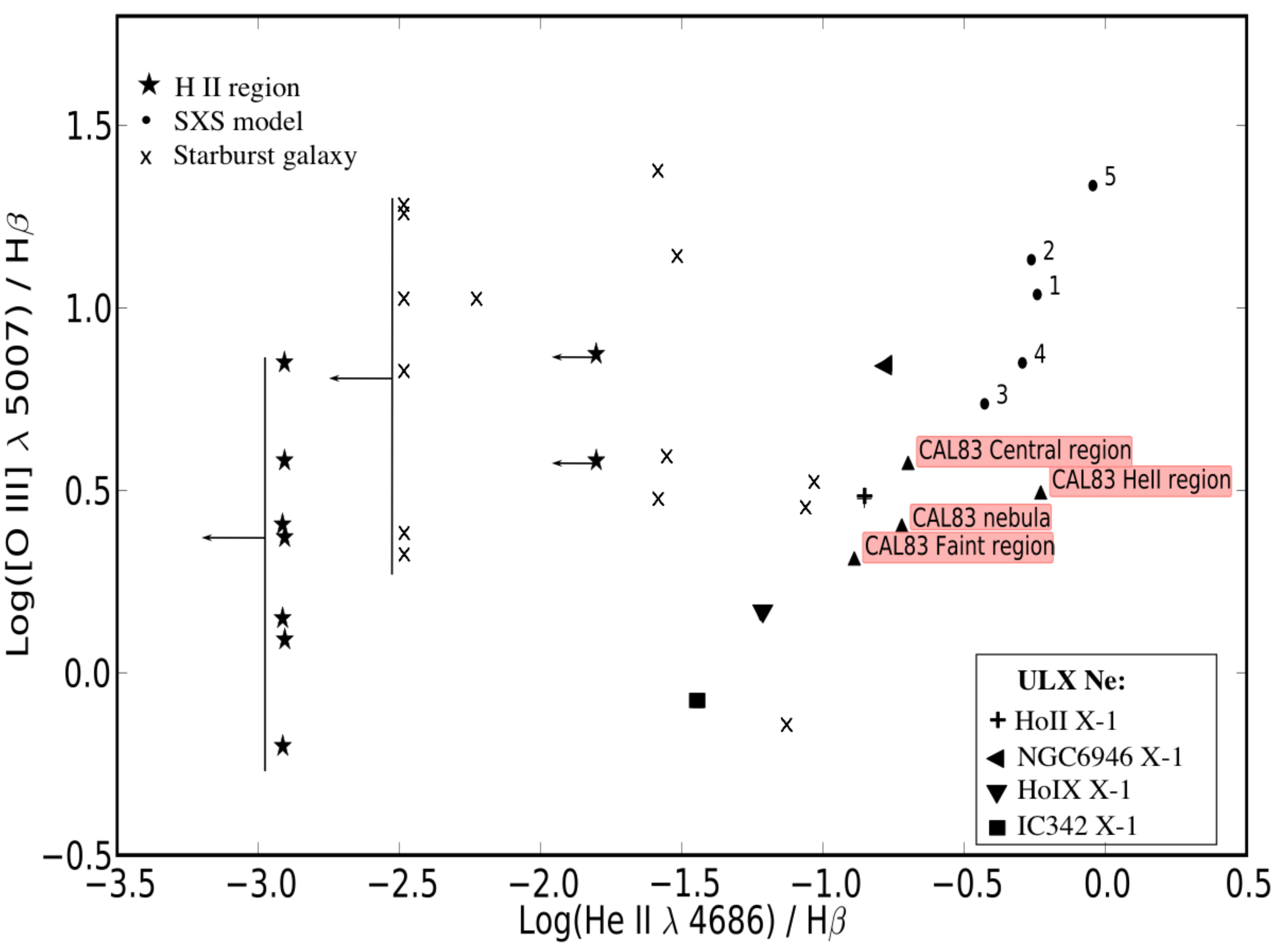}
\caption[The \OthreeB/\Hb\ vs \NtwoB/\Ha\ diagnostic diagram.]{Diagnostic line-ratio diagrams for different regions in the CAL\,83 nebula compared with ultra-luminous X-ray sources (ULX), which have much harder X-ray spectra, as well as star forming galaxies (SFG) and AGNe. ULX data were taken from \citet{Abolmasov2007, Abolmasov2008}. The name for each ULX is given next to the symbols. The positions of different parts of the CAL\,83 nebula are given by filled upright triangles. Solid curves represent the theoretical `maximum starburst' line from \citet{Kewley2001} which distinguishes AGNe from H\,{\scriptsize II} region-like objects (such as SFG). \textit{ Left}: \OthreeB/\Hb\ vs \StwoC/\Ha. \textit{ Right}: \OthreeB/\Hb\ vs \NtwoB/\Ha. \textit{ Bottom}: \OthreeB/\Hb\ vs \Hetwo/\Hb. The filled dots here correspond to our {\sc cloudy} models for SXSs. The numbers near the filled circles indicate the model number from Table\,\ref{Tab:models}. For comparison we have included starburst galaxies and \Htwo\ regions, the data for these objects being taken from the same paper as the SXS models. Arrows represent upper limits; vertical bars with horizontal arrows drawn adjacent to certain starburst galaxies and \Htwo\ regions indicate common sets of upper limits.}\label{Fig:diagnostics}
\end{figure*}

These diagnostic diagrams were designed to distinguish narrow-line active galactic nuclei (AGN) from \Htwo\ region-like objects but can be used for any gaseous emission line-objects. \citet{Chiang1994} suggested another potentially useful diagnostic diagram for SXSs, in particular to intercompare them among \Htwo\ region-like objects. This diagram is based on two important emission lines which should be found in the spectrum of an SXS: \OthreeB\ and \Hetwo. 

We compare our results with the ratios found in ultraluminous X-ray sources (ULXs). The data for these ULXs were taken from the work of \citet{Abolmasov2007, Abolmasov2008}. ULXs can be regarded as the big brothers of SXSs, in the sense that they are harder X-ray sources, but radiate extreme luminosities of $10^{39-41}$ erg s$^{-1}$ in the 0.5--10 keV band, with significant soft components in their X-ray spectra (e.g., \citealt*{Gladstone2009a}). At these high luminosities they exceed the Eddington limit for a $10 M_{\odot}$ black hole, leading to the speculation that they may harbour intermediate-mass black holes (IMBHs; \citealt*{Colbert1999,Miller2004}). However, subsequent analyses support super-Eddington emission scenarios for stellar mass black holes in all but the brightest ULXs (e.g., \citealt{Roberts2007}; \citealt*{Gladstone2009b}). A number of ULXs in nearby galaxies have now been identified with large nebulae (e.g. \citealt*{Pakull2002}; \citealt{Roberts2003, Abolmasov2007}), some of which display characteristics of X-ray ionisation in their central regions. They therefore present an interesting comparator for the nebula around CAL\,83.

We have plotted our results for different areas of the CAL\,83 nebula on the diagrams of Fig. \ref{Fig:diagnostics}. In each we have plotted values for the three regions marked in Fig. \ref{Fig:ratio} (\Stwo/\Ha) and given in Table \ref{Tab:diagnostics}, and one point reflecting the average value over our FoV (excluding the stars). The \HetwoX\ region in the south-east corner of our FoV is interesting because that is where the only nebular \HetwoX\ is found, as well as a peak in H$\alpha$; the faint region in the north-west corner corresponds to the edge of the inner nebula, where the material is dropping in ionisation level; and the central region of our FoV is where \Othree\ and H$\alpha$ are bright. Note that CAL\,83 is the first and only SXS around which an optical ionisation nebula has been detected, so this is the first time an SXS nebula has been placed on these diagnostic plots. 

The CAL\,83 nebula shows characteristics of \Htwo\ region-like objects {\sl and} AGNe. The \NtwoB/\Ha\ ratio characterises our nebula as an \Htwo\ region-like object, while the \StwoC/\Ha\ places it on the side of the AGN-like objects suggesting it has a more extended zone of partially ionised hydrogen than \Htwo\ region-like objects since AGNe are characterised by a harder radiation field. 
The placement of the different regions of CAL 83 shows some spread, but the points always lie in the same general area of the plots as each other. There seems to be no great distinction between this SXS and the ULXs which is interesting since nebulae surrounding ULXs are created by hard X-ray radiation. In the \Hetwo/\Hb\ diagram in Fig.\,\ref{Fig:diagnostics} we include theoretical values we calculate for model SXS nebulae (Sec.\ 5). The models are characterised by high values for the \OthreeB/\Hb\ ratio and substantial \Hetwo\ emission. Our value from the \HetwoX\ region is consistent with the \Hetwo/\Hb\ range suggested by the models but we did not observe the predicted high \OthreeB/\Hb\ flux ratios.



\section{Theoretical Models vs.\ Observations}\label{sect:discussion}

\citet{Chiang1994} created some simple models for SXS nebula, based on a spherical cloud ionised by a hot blackbody. The total abundances and abundance gradients were compared to the little data they had for CAL\,83. However, their plots and tables were taken from models run with Galactic abundances. Since CAL\,83 is in the LMC, we decided to run new models with LMC abundances. 
For this we use the latest version (c10.00) of spectral synthesis code {\sc cloudy} \citep{Ferland1998}. In this section we will compare these models with our observations and discuss the results. Following \citet{Chiang1994}, we have produced 5 simple models for SXSs in which we differentiate between luminosity, temperature and density. For comparison we have also included a model for a classical \Htwo\ region using the physical parameters given in \citet{Evans1986}. The input parameters for the different models can be found in Table \ref{Tab:models}. 

\begin{table}
\caption[Model parameters.]{\label{Tab:models}
Supersoft X-ray Source Model Parameters.}
\begin{center}
\begin{tabular}{lccc} \hline\hline\rule[0mm]{0mm}{3mm}
Model & $T_{\mbox{eff}}$ & $L$ & n  \\
 \hspace{0.4cm}(\#) &($K$) & (\ergs) & (cm$^{-3}$) \\
\hline
1 (standard) & $4\times10^5$ & $1\times10^{38}$ & 10 \\ 
2 & $2\times10^5$ & $1\times10^{38}$ & 10 \\ 
3 & $7\times10^5$ & $1\times10^{38}$ & 10 \\ 
4 & $4\times10^5$ & $1\times10^{37}$ & 10 \\ 
5 & $4\times10^5$ & $1\times10^{38}$ & 1 \\ 
6 (H{\sc ii} R) & $4.6\times10^4$ & $5.1\times10^{38}$ & 10 \\ 
\hline\hline 
\end{tabular}
\end{center}
\end{table}

The ionising source in all models is represented by a point source radiating as a simple blackbody with a given temperature. The point source is placed in the centre of a spherical symmetric and homogeneous gas cloud that has comparable elemental abundances relative to hydrogen of the LMC and were take from \citet{Rolleston2002}. The standard model (model 1) has an X-ray source of temperature $4\times10^5$\,K (${\rm kT}=34$ eV) and a luminosity of $1\times10^{38}$\,\ergs, representative of cited values for SXSs in the literature (\citealt{Trumper1991}; \citealt*{Orio1993}; \citealt*{Hertz1993}; {\citealt*{Kahabka1997}). The hydrogen number density is taken to be 10 hydrogen atoms cm$^{-3}$. The other models are calculated by holding two of the parameters fixed with respect to the standard model and varying the third parameter.

\subsection{Emission line fluxes}
The output of {\sc cloudy} includes intrinsic line intensities and luminosities of the cloud. The intrinsic emission includes all processes that affect the line formation and transfer. This includes collisional processes, fluorescence, line destruction by background opacities such as dust or the Lyman continuum of hydrogen, and recombination. The intrinsic intensities do not include the effects of absorbers or scatters that do not lie within the line-formation region.

\begin{table*}[ht]
\caption[model fluxes.]{\label{Tab:model} Model and observed fluxes.}
\begin{center}
\begin{tabular}{ccccccccc} \hline\hline\rule[0mm]{0mm}{3mm}
$\lambda$ & Ion & Model 1 & Model 2 & Model 3 & Model 4 & Model 5 & Model 6 & Observed \\
\hline
6563 & H I    &   3.08 &   2.93  &  3.22 &  3.10  &   2.97 &  3.04 & 2.46 \\ 
4340 &  H I      &   0.47  &   0.47  &  0.46 &  0.46  &   0.47&  0.46 & 0.34 \\
6678 &  He I     &   0.03  &   0.03  &  0.04  &  0.04  &   0.01 &  0.05 & $<$ 0.08 \\
4686 &  He II    &   0.57  &   0.55  &  0.37  &  0.51  &   0.90 &  0.00 & 0.19 \\
5198 &  [N I]    &   0.60  &   0.11 &  1.15  &  0.81  &   0.01 &  0.00& $< $0.08 \\
5200 &  [N I]    &   0.90 &   0.17  &  1.71  &  1.22 &   0.01 &  0.01 & $<$ 0.08 \\
6548 &  [N II]   &   1.48  &   0.99  &  1.72  &  1.79 &   0.78 &  0.29 & 0.20 \\
6584 &  [N II]   &   4.36  &   2.91  &  5.08  &  5.27  &   2.31&  0.84 & 0.37 \\
6363 &  [O I]    &   1.08  &   0.26  &  1.97  &  1.44  &   0.02 &  0.01 & $<$ 0.09 \\
6300 &  [O I]    &   3.37  &   0.83  &  6.19  &  4.50  &   0.06 &  0.04 & 0.52 \\
4959 &  [O III]  &   3.62  &   4.50  &  1.81 &  2.35  &   7.19 &  0.39 & 0.86 \\
5007 &  [O III]  &  10.88  &  13.55  &  5.46  &  7.07  &  21.64 &  1.18 & 2.50 \\
6716 &  [S II]   &   2.36  &   1.72  &  2.51  &  3.83  &   0.68 &  0.27 & 0.96 \\
6731 &  [S II]   &   1.65  &   1.21  &  1.74  &  2.67  &   0.48 &  0.19 & 0,65 \\
\hline\hline
\end{tabular}
\tablefoot{The \Hb\,-normalised dereddened intensities of the emission lines taken from Table 3 compared to the {\sc cloudy} model intensities. Wavelengths $(\lambda)$ are given in \ang\ and upper limits are indicated with a $<$.}
\end{center}
\end{table*}

Table \ref{Tab:model} gives the values of the emission line fluxes for each of the six model nebulae and our measured values. All emission line intensities in the table are normalised to the corresponding \Hb\ flux to correct for the fact that we only observed about 1/4th of the full nebula. 

Our observed fluxes are those of Table\,3, with 1-$\sigma$ upper limits for [N\,\scriptsize{I}\normalsize]\,$\lambda\lambda5199,5202$, [O\,\scriptsize{I}\normalsize]\,$\lambda6363$ and He\,\scriptsize{I}\normalsize\,$\lambda6678$, as measured from the full FoV spectra using the ELF package in DIPSO \citep{Dipso}. \\
\newline

Unfortunately none of the models is a particularly good match to our observations. For most of the ions the models predict larger line ratios (compared with \Hb) than we observe. The largest discrepancies are: (i) the extremely high \Othree/\Hb\ flux ratio is not observed, where the measured values are typically $\sim$ 1/4 the SXS model predictions and about twice the \Htwo\ region model predictions; (ii) our observed \HetwoX/\Hb\ ratio for the whole nebula is less than half the model predictions, note however that if we just take the \HetwoX\ region (\HetwoX/\Hb\  =  0.59), the ratio does match up; (iii) our observed \Ntwo\ ratios are 5 times to about an order of magnitude lower than the model predictions, and (iv) the \Stwo\ ratios are a factor of $\sim$3 lower than predicted by our standard model.\\
We do not have a ready explanation for these systematic discrepancies. It is possible that the \Othree/\Hb\ ratio in particular is affected by the fact that most of the \Othree\ (as demonstrated by RRM95) is outside our FoV.  Indeed, the spatial inhomogeneity of the CAL\,83 nebula, and the fact that our observations cover only a fraction of the structure, must have implications for our comparison of the simple model to the data that could only be resolved by a spatially complete analysis of the nebula. A more speculative explanation for the discrepancies could be that the CAL\,83 nebula has unusual abundances for the LMC. This could work toward explaining the low nitrogen ratio we observe compared to the sulphur ratio. It could also be that we still do not have a full understanding of the astrophysics of these regions. Further observations of different parts of the nebula will provide us with better constraints on our models and may eventually aid us in better understanding the physics of these unusual nebulae.

\section{Conclusions}

In this paper, we have presented flux maps made from the fitting the emission lines of \OthreeB, \OoneA, \Ha\ and \StwoA\ showing the morphology of these ions in our FoV. The morphology in \OthreeB\ and \Ha\ match those found by RRM95 who imaged the full nebula in these ions. We find the edge of the RRM95 nebula is where low ionisation ions peak, while inside this the ionisation state is higher. 

We have also presented, for the first time, evidence of an \Hetwo\ region around CAL\,83. The \HetwoX\ emission peaks at the position of CAL\,83, but it has a distinctly asymmetrical distribution around the central star. We did not detect any \Heone\ emission.

We estimated an average value for the electron density $n_e$ in our FoV of $\sim$10 cm$^{-3}$ which is consistent with the value found by RRM95.

In addition to the flux maps, we also show flux {\em ratio} maps which characterise zones with different ionisation degrees in our FoV. The flux ratio values of four interesting positions in our FoV were used to place CAL\,83 in the diagnostic diagrams of \citet{Veilleux1987} which is used to distinguish between AGNe and \Htwo\ region-like objects (i.e. star forming galaxies). This has not been done before for a SXS, as CAL\,83 is the only know SXS surrounded by an ionisation nebula. We show that CAL\,83 has characteristics of both AGNe and \Htwo\ region-like objects and does not seem to be distinguishable from ULXs. We have also placed CAL\,83, both observed and modelled, on a plot of \OthreeB/\Hb\ vs \Hetwo/\Hb, developed especially for SXSs by \citet{Chiang1994}. CAL\,83, observed and modelled, differ slightly in the \Hetwo/\Hb\ ratio and more in the \OthreeB/\Hb\ ratio, and is situated with the ULXs and starburst galaxies rather than with the \Htwo\ regions. 

Finally, we have presented a comparison between our observations and model calculations for nebulae surrounding SXSs using the {\tt CLOUDY} ionisation code. Improvement on the models of Rappaport et al. (1994a) was achieved by utilising LMC abundances rather than Galactic. We found that none of the models presented matches our observations completely. Keeping in mind that the observed nebula is not even close to being homogeneous, as is assumed in the modelling, one could argue that the models are perhaps too simplistic. The modelling inconsistencies do not, however, affect the observationally oriented results presented in this paper. Perhaps the discrepancies are telling us something interesting and potentially important about the physical processes that we do not yet understand. To get to the bottom of this, new observations covering the entire CAL\,83 nebula should be compared to more detailed models which perhaps take into account a certain degree of inhomogeneity.



\begin{acknowledgements}

The authors wish to thank the VIMOS support staff at Paranal for taking these service mode observations [program 076.C-0284(C), PI: Roberts].

\end{acknowledgements}

\nocite{Mutchler1997}
\bibliographystyle{aa}
\bibliography{allreferences}
 
\end{document}